\documentclass[12pt]{iopart}

\usepackage{graphicx,iopams,bm}

\def\graphicscale{0.32}
\def\smallgraphicscale{0.32}
\begin{document}

\title[Scaling of entanglement in trapped one-dimensional
       lattice systems]{Scaling of bipartite entanglement in
       one-dimensional lattice systems  with a trapping potential}

\author{Massimo Campostrini and Ettore Vicari}

\address{Dipartimento di Fisica dell'Universit\`a di Pisa and 
  INFN, Sezione di Pisa, Largo Bruno Pontecorvo 2, I-56127 Pisa, Italy} 

\date{May 14, 2010}

\begin{abstract}

  We study the effects of a power-law trapping potential on the scaling
  behaviour of the entanglement at the quantum critical point of
  one-dimensional (1D) lattice particle systems.  We compute bipartite von
  Neumann and R\'enyi entropies in the presence of the trap, and analyze their
  scaling behaviour with increasing the trap size.
  
  As a theoretical laboratory, we consider the quantum XY chain in an external
  transverse field acting as a trap for the spinless fermions of its quadratic
  Hamiltonian representation.  We then investigate confined particle systems
  described by the 1D Bose-Hubbard model in the superfluid phase (around the
  center of the trap).
  
  In both cases conformal field theory predicts logarithmically divergent
  bipartite entanglement entropies for the homogeneous systems without trap.
  The presence of the trapping potential breaks conformal invariance,
  affecting the critical behaviour of the homogeneous system. Our results show
  that the bipartite entanglement entropies diverge logarithmically with
  increasing the trap size, and present notable scaling behaviours in the
  trap-size scaling limit.

\end{abstract}



\section{Introduction}
\label{intro}

The spatial entanglement of one-dimensional (1D) lattice systems near their
quantum critical point is a physical issue which has attracted much interest,
see, e.g., Ref.~\cite{CCD-09}.  Many studies have been devoted to quantify it,
by computing von Neumann or R\'enyi entanglement entropies of the reduced
density matrix of a subsystem.  In a homogeneous system of length $L$ whose
quantum critical behaviour is described by a two-dimensional (2D) conformal
field theory (CFT), the bipartite entanglement entropy $S(l_A,L)$ increases
logarithmically~\cite{HLW-94,VLRK-03,CC-04,CC-09} with increasing the size
$l_A$ of the subsystem at the critical point, i.e., $S\sim \ln l_A$, or with
increasing the spatial length scale $\xi$ when approaching the critical point,
i.e., $S\sim \ln \xi$ (if $1\ll\xi\ll l_A$).  In both cases the coefficients
of the logarithms are related to the central charge of the corresponding CFT.

Quantum lattice particle systems are realized in experiments of cold atoms in
optical lattices of any effective spatial dimension $d\le 3$, which allow to
investigate the interesting interplay between quantum and statistical
behaviours in particle systems, see, e.g.,
Refs.~\cite{BDZ-08,GBMHS-02,SMSKE-04,PWMMFCSHB-04,KWW-04,KWW-05,HSBBD-06,HSBBD-07,FWMGB-06,
  SPP-07,SPP-08,CFFFI-09}. An important feature of these experiments is the
presence of a confining potential which traps the particles within a limited
spatial region of the optical lattice created by laser-induced standing waves.
The theoretical framework~\cite{JBCGZ-98} is based on the Bose-Hubbard (BH)
model~\cite{FWGF-89} in the presence of a confining potential coupled to the
particle density, i.e.,
\begin{equation}
H_{\rm BH} = -{J\over 2}
\sum_{\langle ij\rangle}(b_i^\dagger b_j+b_j^\dagger b_i) 
+\, {U\over2} \sum_i n_i(n_i-1)+
\mu \sum_i n_i + \sum_i V(r_i) n_i ,\quad
\label{bhm}
\end{equation}
where $\langle ij\rangle$ is the set of nearest-neighbour sites, $n_i\equiv
b_i^\dagger b_i$ is the particle density operator.  We consider a power-law
trapping potential
\begin{equation}
V(r) = v^p r^p \equiv (r/l)^p,
\label{potential}
\end{equation}
where $r\equiv |\vec{x}|$, $v$ and $p$ are positive constants and $l\equiv
1/v$ is the trap size.  Experiments are usually set up with a harmonic
potential, i.e., $p=2$.  Far from the origin the potential $V(r)$ diverges,
therefore $\langle n_i\rangle$ vanishes and the particles are trapped.  The
inhomogeneity due to the trapping potential strongly affects the phenomenology
of quantum transitions in homogeneous systems.  The correlation functions
develop critical behaviours only in the limit of large trap size.  This
critical behaviour can be described in the framework of the trap-size scaling
(TSS) theory~\cite{CV-10,CV-09}.

In this paper we focus on the quantum entanglement in lattice particle systems
confined by an external potential, such as the BH model (\ref{bhm}).  We
investigate how the scaling behaviour of the entanglement entropies determined
by CFT changes in the presence of the inhomogeneity induced by the confining
potential.

As a theoretical laboratory we consider the quantum XY chain in an external
space-dependent transverse field, acting as a trap for the spinless fermions
of its quadratic Hamiltonian representation.  This model presents a quantum
critical behaviour belonging to the 2D Ising universality class, thus
described by a CFT with central charge $c=1/2$.  In the presence of the trap
and for large trap sizes, the quantum critical behaviour shows
TSS~\cite{CV-10,CV-09}.  The presence of the trap induces a length scale $\xi$
at the critical point, which scales nontrivially with increasing the trap
size, i.e., as $\xi\sim l^{\theta}$ with a trap exponent $\theta=p/(p+1)$.

We then consider the 1D BH model (\ref{bhm}) in its hard-core limit, i.e.,
$U\to\infty$, and study the bipartite entanglement entropy in the gapless
superfluid phase, whose continuum limit is described by a 2D massless
bosonic field theory, thus a CFT with $c=1$. In this case the trap-size
dependence is made  more complicated by new features~\cite{CV-10-2}:
(i) a multiscaling phenomenon, i.e., the 
existence of different length scales which diverge with distinct power laws in
the TSS; (ii) the presence of level crossings of the lowest states at finite
trap size, which gives rise to a peculiar modulation of the amplitudes of the
asymptotic power-law behaviours.

We analyze the scaling behaviour of bipartite von Neumann and R\'enyi
entanglement entropies in these 1D lattice models. We consider chains with
even sites $L$ and open boundary conditions, and a trap of size $l$ centered
between the middle sites of the chain.  We divide the chain in two connected
parts of length $l_A$ and $L-l_A$, and consider their von Neumann entropy
\begin{equation}
S_1(l_A;L)=S_1(L-l_A;L) = -{\rm Tr}[\rho_A \ln \rho_A]
\label{vNen}
\end{equation} 
and their R\'enyi entropies
\begin{equation}
S_\alpha(l_A;L) = S_\alpha(L-l_A;L) = 
{1\over 1-\alpha} \ln {\rm Tr} \rho_A^\alpha
\label{renyientropies}
\end{equation}
where $\rho_A$ is the reduced density matrix of one of the two subsystems.
R\'enyi entropies are useful because they provide information on the spectrum
of $\rho_A$ \cite{CL-08}.  The von Neumann entropy (\ref{vNen}) is obtained by
taking the limit $\alpha\to 1$ in (\ref{renyientropies}).  In the presence of
a trap of size $l$, the bipartite entanglement entropies of subsystems of size
$l_A=L/2-x$ have a finite $L\to\infty$ limit, depending on $x$ and $l$.  We
study their scaling behaviour in the large-$l$ limit.  Note that the
$p\to\infty$ limit of the confining potential becomes equivalent to a
homogeneous chain of size $L=2l$ with open boundary conditions (more
precisely, to a chain with even $L=2\lfloor l\rfloor$ sites).  Thus, in the
$p\to\infty$ limit we must recover the entanglement entropies of homogeneous
systems with open boundary conditions, which are determined by CFT.

In our numerical study we exploit the quadratic spinless fermion
representations of the XY chain and the 1D hard-core BH model, which allow us
to perform computations for very large systems, since they only require the
diagonalization of a $L\times L$ matrix where $L$ is the number of lattice
sites.  We compute the entanglement entropies using the method outlined in
Refs.~\cite{Peschel-03,LR-09}, which exploits the quadratic fermionic
representation and can be also applied in the presence of the trapping
potential.  We obtain numerical results for chains of size $L$ and open
boundary conditions, with a trap of size $l$ centered between the middle sites
of the chain; we choose $L$ large enough to have negligible finite-$L$
effects; we are able to obtain very accurate results for $l$ up to
$\Or(10^4)$.

Our results show that the bipartite entanglement entropies at the critical
point diverge logarithmically with increasing the trap size. Moreover, they
present notable scaling behaviours in the trap-size scaling limit.

We mention that the bipartite entanglement entropies in the XY and XX model
with gradients, i.e., in the presence of a linear external field, have been
recently investigated in Ref.~\cite{EIP-09}.

The paper is organized as follows. In Sec.~\ref{sec1} we presents results for
the quantum XY chain in an external space-dependent transverse field.
Sec.~\ref{sec2} is dedicated to the 1D hard-core BH model.  In
Sec.~\ref{conclusions} we draw our conclusions.

\section{Bipartite entanglement in the XY chain with a space-dependent
  transverse field}
\label{sec1}

The quantum XY chain in a transverse field is a standard theoretical
laboratory for issues related to quantum transitions, see, e.g.,
Ref.~\cite{Sachdev-book}. An inhomogeneity analogous to the one
arising from a trapping potential in particle systems can be achieved by
considering a space-dependent transverse field, i.e.,
\begin{equation}
H_{\rm XY} = -  \sum_i {1\over 2} [ (1+\gamma) \sigma^x_i \sigma^x_{i+1}
+ (1-\gamma) \sigma^y_i \sigma^y_{i+1}] 
- \mu \sigma^z_i  - V(x_i) \sigma^z_i,
\label{Isc}
\end{equation}
where $\sigma^i$ are the Pauli matrices, $\gamma\ne 0$ and $V(x) = (|x|/l)^p$,
cf.\ (\ref{potential}).  The XY chain can be mapped into a quadratic
Hamiltonian of spinless fermions by a Jordan-Wigner transformation, obtaining
\begin{equation}
\eqalign{H = \sum [ c_{i}^\dagger A_{ij} c_{j}  + {1\over 2}
(c_{i}^\dagger B_{ij} c_{j}^\dagger + {\rm h.c.})], \label{sfi} \\
A_{ij} = 2 \delta_{ij} - \delta_{i+1,j} - \delta_{i,j+1}
+ 2 Q(x_i) \delta_{ij},  \\
B_{ij} = - \gamma \left( \delta_{i+1,j} - \delta_{i,j+1} \right), \\
Q(x)= \bar{\mu}+V(x),\quad \bar{\mu}\equiv \mu-1.}
\end{equation}
In this picture $\mu$ plays the role of {\em chemical potential} for the
fermion $c$-{\em particles}, and the space-dependent field $V(x)$ acts as a
{\em trap} for the $c$-particles, making their {\em local density} $\langle
n_i\rangle\equiv \langle c^\dagger_i c_i \rangle$ vanish at large distance.
In the following, $l$ will be considered as the {\em trap size}.

In the absence of the trap, the model undergoes a quantum transition at
$\bar{\mu}\equiv \mu-1=0$ in the 2D Ising universality
class, separating a quantum paramagnetic phase for $\bar{\mu}>0$ from a
quantum ferromagnetic phase for $\bar{\mu}<0$.  Around $\bar{\mu}=0$, the
quantum critical behaviour shows a diverging length scale, $\xi\sim
|\bar{\mu}|^{-\nu}$, and a vanishing energy scale, $\Delta\sim \xi^{-z}$,
where $z$ and $\nu$ are universal critical exponents: $z=1$ and $\nu\equiv
1/y_\mu=1$ ($y_\mu$ is the RG dimension of $\mu$).

The presence of the external field $V(x)$ gives rise to a space inhomogeneity,
thus affecting the critical behaviour of the homogeneous system, which is only
recovered in the limit of large trap size.  In this limit the critical
behaviour can be described in the framework of the trap-size scaling (TSS)
theory~\cite{CV-10}.  The TSS Ansatz for the asymptotic behaviour of the
free-energy density in the presence of a confining potential (\ref{potential})
is
\begin{equation}
F(\mu,T,l,x) = l^{-\theta (1+z)} 
{\cal F}(\bar{\mu} l^{\theta/\nu},Tl^{\theta z},xl^{-\theta}),
\label{freee}
\end{equation}
where $x$ is the distance from the middle of the trap, and $\theta$ is the
trap-size exponent which determines how the length scale of the critical modes
at the critical point diverges with increasing trap size, i.e., $\xi\sim
l^{\theta}$. RG arguments and analytic calculations show that~\cite{CV-10} 
\begin{equation}
\theta=p/(p+1).
\label{thetavalXY}
\end{equation}
The TSS limit can be analytically derived within the quadratic spinless
fermion representation (\ref{sfi}), obtaining a Schr\"odinger-like equation
for the lowest states, after introducing the rescaled spatial variable $X
\equiv \gamma^{-\theta/p} l^{-\theta} x$, and the rescaled deviation from the
critical point $\mu_r \equiv \gamma^{-\theta} l^{\theta} \bar{\mu}$ (see
Ref.~\cite{CV-10} for details).  The $\gamma$ dependence gets absorbed by such
rescalings leading to a universal TSS limit. Therefore at the quantum critical
point $\bar{\mu}=0$ and $T=0$, any length scale related to the critical modes
must have the asymptotic behaviour
\begin{equation}
\xi = a \gamma^{\theta/p} l^{\theta} \left[ 1 + \Or(l^{-\theta})\right]
\label{anyxisc}
\end{equation} 
where the amplitude $a$ depends on the power $p$ of the potential, but not on
$\gamma$. The power law of the scaling corrections is inferred from the
analysis of the corrections to the continuum Schr\"odinger-like equation
describing the TSS limit.

We want to study the quantum entanglement of the ground-state in the presence
of the external space-dependent potential (\ref{potential}). For this purpose
we analyze the trap-size dependence of the bipartite von Neumann and R\'enyi
entanglement entropies, defined in equations (\ref{vNen}) and
(\ref{renyientropies}), in the framework of the TSS theory.

Before presenting the results in the presence of the trap, let us recall some
established results for the homogeneous XY chain with open boundary
conditions, which represents the formal $p\to\infty$ limit of the model with
the confining potential.  They can be obtained by exploiting the CFT that
describes its continuum limit.  Setting
\begin{equation}
l_A=L/2-x
\label{ladef}
\end{equation}
(we consider a chain with
an even number of sites $L$), the R\'enyi entropies at the critical
point $\bar{\mu}=0$ behave as~\cite{CC-04}
\begin{equation}
S_\alpha(L/2-x;L) \approx C_\alpha
\left[ \ln L + \ln\cos\left({\pi x\over L}\right) + e_\alpha \right] 
\label{ccfo}
\end{equation}
where 
\begin{equation}
C_\alpha\equiv c\,{1+\alpha^{-1}\over 12} 
\label{Calpha}
\end{equation}
and $c$ is the central charge: $c=1/2$ for the 2D Ising universality
class.  The constant $e_\alpha$ can be derived using the results of
Refs.~\cite{CC-04,JK-04,IJ-08,CV-10}, obtaining
\begin{equation}
\eqalign{
 e_\alpha = \ln \gamma + \ln (8/\pi) + y_\alpha, \\
 y_\alpha = \int_0^\infty \frac{\rmd t}t 
 \left[{6\over 1-\alpha^{-2}} \left(\frac1{\alpha\sinh t/\alpha} -
      \frac1{\sinh t}\right) \frac1{\sinh t}- \rme^{-2t}\right].}
\label{egamma}
\end{equation}
The behaviour of the von Neumann entropy can be obtained by taking the limit
$\alpha\to 1$ of (\ref{ccfo}).  Corrections to the asymptotic behaviour
(\ref{ccfo}) are expected to be $\Or(L^{-1/\alpha})$ at fixed
$x/L$~\cite{CC-10,CCEN-10}.  They become logarithmic for $\alpha\to\infty$.
In the case of the von Neumann entropy, i.e., $\alpha\to 1$, corrections are
$\Or(L^{-1})$.\footnote{It is worth mentioning that, using the
  relation~\cite{IJ-08} $S_\alpha(l_A;L)|_{\gamma=1}={1\over 2}
  S_\alpha(2l_A;2L)|_{\gamma=0}$ and results from
  Refs.~\cite{LSCA-06,CV-10-2}, we obtain $S_1(L/2;L) = {1\over 12} \left[\ln
    L + e_1 + (2-3\pi)/(4L) +\Or(1/L^2)\right]$ for $\gamma=1$.}

In the presence of the trap of size $l$, the dependence on $L$ rapidly
disappears for $L$ sufficiently larger than the trap size $l$.  We define
entanglement length scales $\xi_\alpha$ from the half-lattice entanglement
entropies.  We write (at $\bar{\mu}=0$)
\begin{equation}
S_{\alpha,1/2}\equiv\lim_{L\to\infty} S_\alpha(L/2;L) \equiv 
C_\alpha \left( \ln \xi_\alpha + e_\alpha\right) ,
\label{ccfot}
\end{equation}
where $e_\alpha$ is the constant defined in (\ref{egamma}). We
identify the argument 
of the logarithm as the entanglement length scale $\xi_\alpha$, which will
generally depend on $\alpha$ and on the power $p$ of the potential
(\ref{potential}).  However, by construction, their $p\to\infty$ limit is
$\xi_\alpha\to 2l=L$ for any $\alpha$.  Since the entanglement length scales
are clearly related to the critical modes, they are expected to have the
asymptotic behaviour (\ref{anyxisc}) at the critical point, although their
scaling corrections may turn out to be different, depending on $\alpha$.  

The approach to their asymptotic behaviour $\xi_\alpha\sim l^{\theta}$ may be
guessed by recalling that the $p\to\infty$ limit is equivalent to a
homogeneous chain of size $L=2l$ with open boundary conditions, where the
corrections to the asymptotic behaviour (\ref{ccfo}) of the R\'enyi
entanglement entropies are~\cite{CCEN-10} $\Or(L^{-1/\alpha})$. Then, the
correspondence $L\to \xi\sim l^{\theta}$ suggests the presence of
$\Or(\xi^{-1/\alpha})$, thus $\Or(l^{-\theta/\alpha})$, corrections.  In the
case of the length scale derived from the von Neumann entropy, i.e., for
$\alpha\to 1$, this argument leads to the same scaling corrections of equation
\eref{anyxisc}, which are expected for any length scale defined from standard
correlation functions, such as $G_{ij}\equiv \langle \sigma^x_i
\sigma^x_j\rangle$.

\begin{figure}[tbp]
\begin{center}
\includegraphics*[scale=\graphicscale]{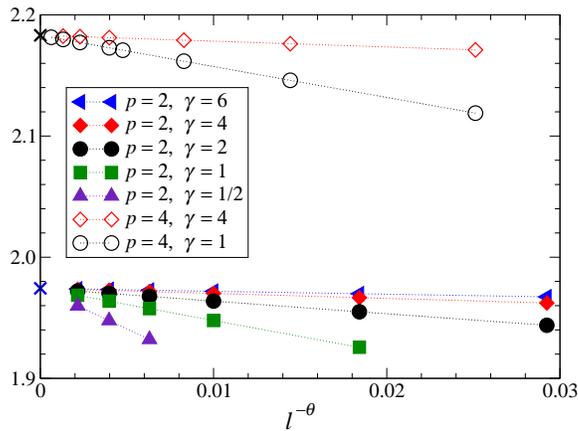}
\end{center}
\vskip-5mm
\caption{
  The ratio $\xi_1/(\gamma^{\theta/p}l^{\theta})$ vs $l^{-\theta}$ for
  $p=2,\,4$, several values of $\gamma$, and trap sizes in the range
  $10^2\le l \le 10^4$.  The lines are drawn to guide the eyes. The crosses
  along the $y$-axis represent the $\gamma$-independent $l\to\infty$
  extrapolations, see text.  }
\label{xi1fig}
\end{figure}

The scaling (\ref{anyxisc}) is fully supported by numerical results for the
half-lattice von Neumann entropy.~\footnote{The effects of the finite size $L$
  of the chain are under complete control in our numerical calculations at
  fixed trap size. This is easily achieved by comparing data at fixed trap
  size for increasing chain size, due to the fast convergence to the
  infinite-chain limit, see also later.  We choose $L$ large enough to have
  negligible finite-$L$ effects, so that the results that we present can be
  effectively considered as infinite-chain results with great accuracy.
  Therefore, the systematic error related to the finite size of the chain is
  totally negligible in our analyses.}  Figure \ref{xi1fig} shows
the results of $\xi_1$ for $p=2$ and $p=4$, several values of $\gamma$, and
for trap sizes in the range $10^2\le l \le 10^4$.  In order to estimate the
amplitude $a_1$ of the asymptotic behaviour
\begin{equation}
\xi_1 =  a_1 \gamma^{\theta/p}l^{\theta}\left[ 1 + b_{11} l^{-\theta} +
  ...\right],
\label{xisca}
\end{equation}
we fit the data of the ratio $\xi_1/(\gamma^{\theta/p}l^\theta)$ to the Ansatz
$a_1 + b_{1,1} l^{-\theta} + b_{1,2} l^{-2\theta}$, which turns out to be
optimal, and check the stability of the results by increasing the minimum
value of $l$ allowed in the fit. The analysis of the $p=2$ data gives
$a_1\approx 1.97431$ for all $\gamma=6,4,2,1,1/2$,\footnote{%
Here and in the following the uncertainty on the numerical estimates is
at most one on the last reported figure.}
confirming its
independence of $\gamma$, and $b_{1,1}=\bar{b}_{1,1} \gamma^{4/3}$ with
$\bar{b}_{11}\approx -2.677$ independent of $\gamma$.  For $p=4$ we obtain
$a_1\approx 2.18316$ for both $\gamma=4,\,1$.

In the case of the entanglement length scales $\xi_\alpha$ defined from the
R\'enyi entropies, we generally expect $\xi_\alpha\sim l^\theta$
asymptotically, with $\Or(l^{-\theta/\alpha})$ corrections.  Assuming
universality of the TSS limit, we expect that the ratios $\xi_\alpha/\xi_1$
approach universal values $R_\alpha$,
\begin{equation}
\xi_\alpha/\xi_1 = R_\alpha + \Or(l^{-\theta/\alpha})
\label{rxia}
\end{equation}
and therefore
\begin{equation}
\xi_\alpha = a_\alpha \gamma^{\theta/p} l^{\theta} \left[ 1+ 
b_{\alpha,1}(\gamma) l^{-\theta/\alpha} + 
b_{\alpha,2}(\gamma) l^{-2\theta/\alpha} + ...\right],
\label{xia}
\end{equation}
with $a_\alpha$ dependent on $p$ but not on $\gamma$.  Note that $R_\alpha\to
1$ and $a_\alpha\to 2$ for $p\to \infty$.

\begin{figure}[tbp]
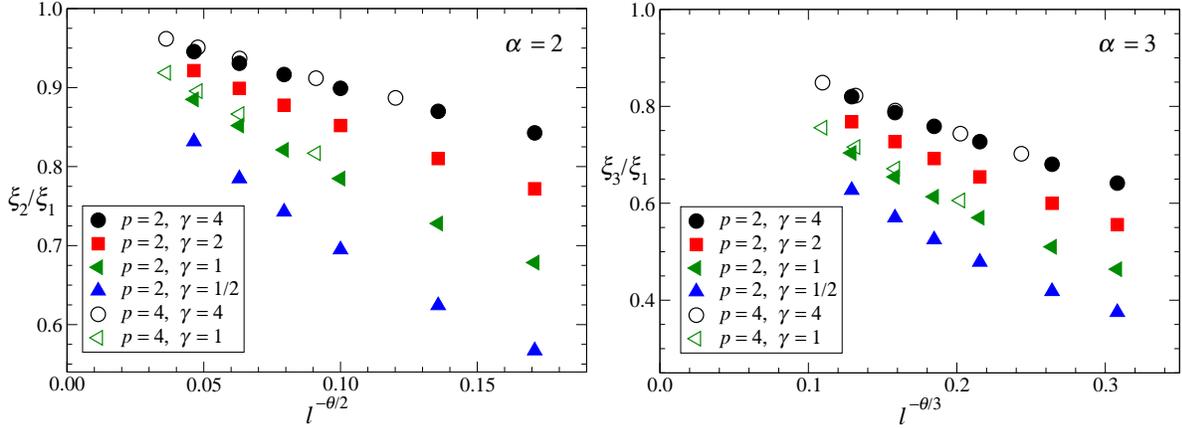

\begin{center}
\hbox{\includegraphics*[scale=\smallgraphicscale]{fig2a.eps}
\includegraphics*[scale=\smallgraphicscale]{fig2b.eps}}
\end{center}
\vskip-10mm
\caption{Ratios 
  $\xi_\alpha/\xi_1$ of entanglement length scales vs.\
  $l^{-\theta/\alpha}$a in the presence of an
  external potential with $p=2,\,4$,  for
  $\alpha=2,\,3$, several values of $\gamma$, and
  trap sizes in the range $10^2\le l \le 10^4$.  }
\label{xiratios}
\end{figure}

This scenario is confirmed by numerical results for the ratios
$\xi_\alpha/\xi_1$, see figure \ref{xiratios} where we show results for
$p=2,\,4$ and $\alpha=2,\,3$.  They fit the general behaviour
\begin{equation}
\xi_\alpha/\xi_1 = R_\alpha + \sum_{j=1} c_j l^{-j\theta/\alpha},
\label{fitansatz}
\end{equation}
and provide consistent values of $R_\alpha$ for different values of $\gamma$,
supporting universality. For example, we find
$R_2\approx 0.9889$, $R_3\approx 0.987$ for $p=2$, 
and $R_2\approx 0.9970$ and $R_3\approx 0.996$ for $p=4$, 
which are slightly smaller than one, and appear to approach one
with increasing $p$.\footnote{To get an idea of the accuracy of the
  universality check, we mention that, using the six available data in the
  range $10^2\le l\le 10^4$ to fix the six parameters of the polynomial
  $R_\alpha + \sum_{j=1}^5 c_j \rme^{-j\theta/\alpha}$, we obtain
  $R_2=0.98892,\,0.98892,\,0.98892, \,0.98889$ and
  $R_3=0.9867,\,0.9867,\,0.9868,\,0.9869$ for $\gamma=4,\,2,\,1,\,1/2$
  respectively.} 

Moreover, the results for $p=2$ indicate that the amplitudes $b_{\alpha,1}$ of
the leading $\Or(l^{-\theta/\alpha})$ scaling corrections in equation
\eref{xia} behave as $b_{\alpha,1}=\bar{b}_{\alpha,1} \gamma^{-\kappa/\alpha}$
where $\kappa = 4/3$ and $\bar{b}_{\alpha,1}$ is independent of $\gamma$, with
great numerical accuracy.  Analogous results are found for $p=4$, but with
$\kappa=6/5$.  Taking into account that $b_{\alpha,1}\sim \gamma^{-1/\alpha}$
for $p\to\infty$ (as inferred by numerical computations), we may guess
$\kappa=(p+2)/(p+1)$.  Therefore, we have that the ratios
$b_{{\alpha},1}^{\alpha}/b_{1,1}$ are independent of $\gamma$.  Note that this
also implies that the leading $\Or(l^{-\theta/\alpha})$ scaling corrections
get suppressed with increasing $\gamma$ for any $\alpha$, including $\alpha\to
1$.

Let us now study the spatial dependence of the entanglement entropy in the
presence of the trap.  For this purpose we define the function
\begin{equation}
  s_\alpha(x)\equiv 
  C_\alpha^{-1} \lim_{L\to\infty} \left[ S_\alpha(L/2-x;L) - S_\alpha(L/2;L) 
\right] 
 \label{xdepstrap} 
\end{equation}
This limit is finite and depends only on $x$ and the trap size $l$.  We expect
that in the TSS limit
\begin{equation}
  s_\alpha(x)\approx f_\alpha(X), \qquad X \equiv x/\xi_1,
 \label{xdepssca} 
\end{equation}
i.e., they are scaling functions of $X$.  Note that the distance of the site
$L/2-x$ from the center of the trap is actually $y=x+1/2$, but the effects of
this difference disappears in the TSS limit $\xi_1\to\infty$ at fixed $X\equiv
y/\xi_1$, as $\Or(\xi_1^{-1})$.  The large-$p$ limits of $f_\alpha$ coincide,
indeed, since $\xi_1\to L$, (\ref{ccfo}) implies
\begin{equation}
\lim_{p\to\infty} f_\alpha(X) = \ln \cos(\pi X)
\label{largeplimit}
\end{equation}
with $|X| <1/2$.  
Corrections to scaling are again expected to be $\Or(l^{-\theta/\alpha})$.

\begin{figure}[tbp]
\begin{center}
\includegraphics*[scale=\graphicscale]{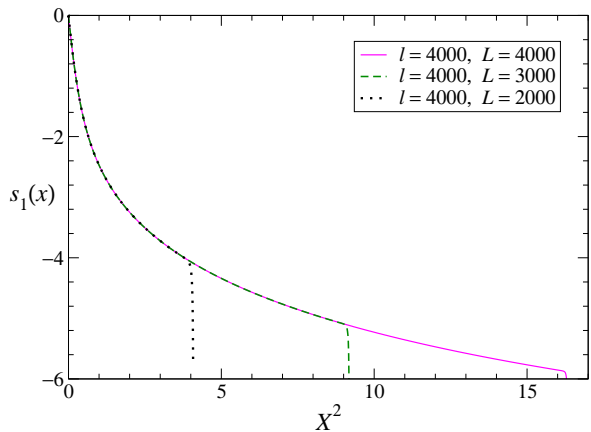}
\end{center}
\vskip-5mm
\caption{
  $s_1(x)$, cf.\ (\ref{xdepstrap}), vs.\ $X^2$, for $\gamma=1$, $p=2$,
  $l=4000$ and a few values of the size of the chain $L$.}
\label{s1L}
\end{figure}

In figure \ref{s1L} we show data for $p=2$, $\gamma=1$, $l=4000$ and a few
values of the size of the chain $L$.  They show the effects of the finite size
$L$ of the chain, and that accurate results for larger and larger $X$ require
larger and larger $L$. However, such effects can be carefully controlled, so
that the results that we will present can be effectively considered as
infinite-chain results.~\footnote{The comparison of data in Fig.~\ref{s1L}
  shows that we can obtain very accurate results around the middle of the trap
  even without requiring $l\ll L$, as one might naively expect. This is
  essentially related to the fact that the relevant length scale of the
  critical correlations behaves as $\xi\sim l^{\theta}$ and $\theta<1$, so
  that we can have $\xi\ll L$ even when $l\approx L$. Note that, with
  increasing the power $p$ of the confining potential, $\theta$ increases and
  approaches one, thus larger chain sizes are required at fixed trap size (we
  recall that for $p\to\infty$ we have the formal correspondence $L=2l$).}  We
used chains of size up to $L=4000$ for $\gamma=1$ and up to $L=6000$ for
$\gamma=4$, which ensure negligible finite-$L$ effects at the trap size
considered.

\begin{figure}[tbp]
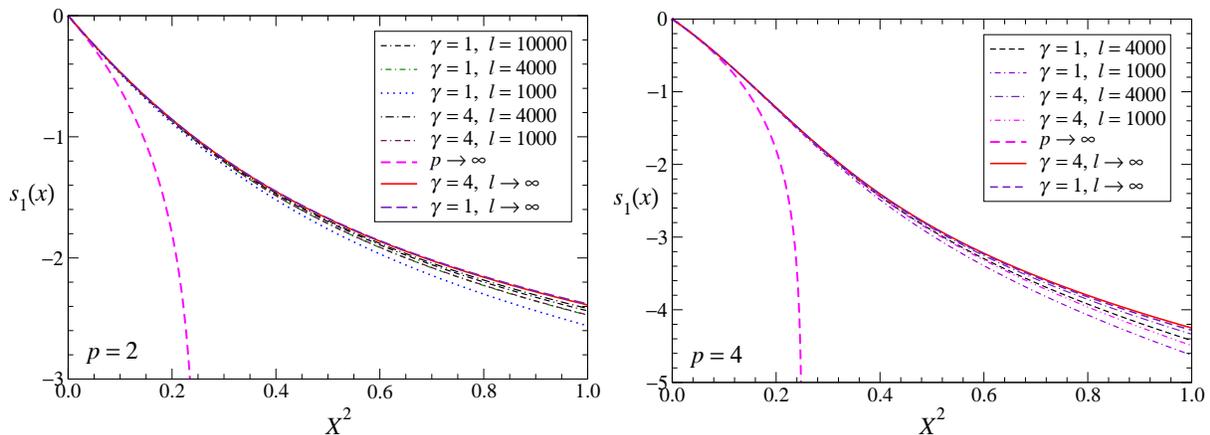

\begin{center}
\hbox{\includegraphics*[scale=\graphicscale]{fig4a.eps}
\includegraphics*[scale=\graphicscale]{fig4b.eps}}
\end{center}
\vskip-10mm
\caption{ 
  The function  $s_1(x)$, cf.\ (\ref{xdepstrap}), vs.\ $X^2$, for $p=2,\,4$,
  $\gamma=1,\,4$, and a few values of $l$.  Curves for
  $l\to\infty$ are obtained by using the data for the three largest
  values of $l$,
  extrapolating at fixed $X$ using the second-order polynomial $\sum_{n=0}^2
  c_n l^{-n\theta}$.  The extrapolated curves of data at $\gamma=1,\,4$ can be
  hardly distinguished, supporting the universality of the TSS.  For
  comparison we also show the $p\to\infty$ limit (\ref{largeplimit}).}
\label{s1xis}
\end{figure}

\begin{figure}[tbp]
\begin{center}
\includegraphics*[scale=\graphicscale]{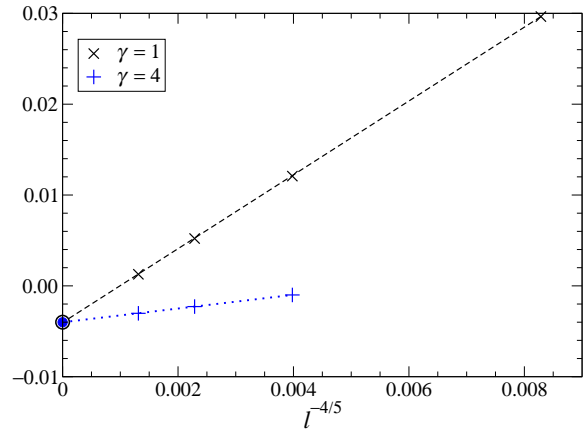}
\end{center}
\vskip-5mm
\caption{
  Results for $c_2-c_2|_{p\to\infty}=c_2+\pi^2/2$, cf.\ (\ref{smallx}), for
  $p=4$, using data for $\gamma=1,\,4$, versus $l^{-\theta}$.  The crosses
  indicate the estimates of $c_2$ from polynomial fits of $s_1(x)$ at small
  $x$, which are expected to converge to its
  $l\to\infty$ with $\Or(l^{-\theta})$ corrections. The lines show linear fits
  of these data to $c_2+a_\gamma l^{-\theta}$. The blobs along the
  $y$-axis show the estimates of $c_2$ from the fits of the $l\to\infty$
  extrapolated curves of $s_1(x)$ for $\gamma=1,\,4$. The overall agreement
  with a universal value of $c_2$ is very accurate, leading to the estimate
  $c_2-c_2|_{p\to\infty}\approx -0.0040$.  }
\label{c2extrap4}
\end{figure}

The numerical results for the von Neumann entropy, reported in
figure \ref{s1xis} for $p=2$ and $p=4$, show that $s_1(x)$ approaches
a scaling function of $X\equiv x/\xi_1$ in the TSS limit (more
precisely, a function of $X^2$, since it 
is an even function of $X$), which is universal, i.e., independent of
$\gamma$.  In figure \ref{s1xis} we also
show extrapolations to $l\to\infty$ assuming $\Or(l^{-\theta})$ corrections.
They are consistent with a universal TSS limit independent of
$\gamma$.  With increasing $p$, the curves approach the $p\to\infty$ limit
(\ref{largeplimit}).  The approach is particularly fast at small $X$, where
we find
\begin{equation}
f_1(X) = c_2 X^2 + \Or(X^4)
\label{smallx}
\end{equation}
with $c_2\approx -4.988$ for $p=2$ and $c_2\approx -4.939$ for $p=4$ (see
figure \ref{c2extrap4}), to be compared with 
$c_2=-\pi^2/2\approx -4.93480$ for $p\to\infty$, 
obtained by expanding equation \eref{largeplimit}. 

\begin{figure}[tbp]
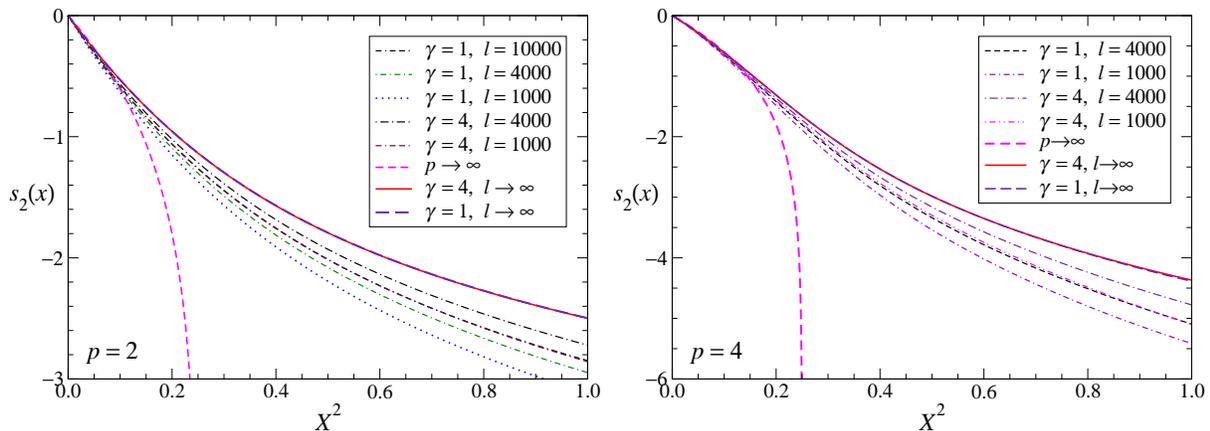

\begin{center}
\hbox{\includegraphics*[scale=\graphicscale]{fig6a.eps}
\includegraphics*[scale=\graphicscale]{fig6b.eps}}
\end{center}
\vskip-10mm
\caption{ 
  The function $s_2(x)$, cf.\ (\ref{xdepstrap}), vs.\ $X^2$, for $p=2,\,4$,
  $\gamma=1,\,4$, and a few values of $l$.
  Curves for $l\to\infty$ are obtained by using the data for the three
  largest values of $l$, extrapolating at fixed $X$ using the
  second-order polynomial 
  $\sum_{n=0}^2 c_n l^{-n\theta/2}$.  The extrapolated curves of data at
  $\gamma=1,\,4$ can be hardly distinguished, supporting the universality of
  the TSS.  For comparison we also show the $p\to\infty$ limit
  (\ref{largeplimit}).}
\label{s2p24}
\end{figure}

\begin{figure}[tbp]
\begin{center}
\includegraphics*[scale=\graphicscale]{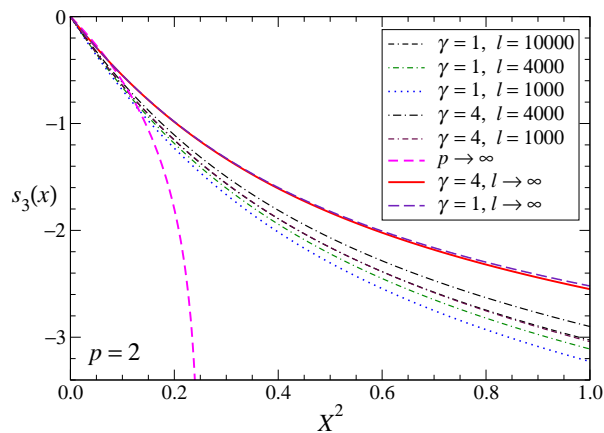}
\end{center}
\vskip-5mm
\caption{ 
  The function $s_3(x)$, cf.\ (\ref{xdepstrap}), vs.\ $X^2$, for $p=2$,
  $\gamma=1,\,4$, and a few values of $l$.  Curves for $l\to\infty$ are
  obtained by using the data for the largest values of $l$,
  extrapolating at fixed $X$ using the polynomials 
  $\sum_{n=0}^{\bar n} c_n l^{-n\theta/3}$ with ${\bar n}=2$ for
  $\gamma=4$ and ${\bar n}=3$ for $\gamma=1$.
  For comparison we also show the $p\to\infty$ limit
  (\ref{largeplimit}).}
\label{s3p24}
\end{figure}

Analogous results are obtained from the R\'enyi entropies, see figures
\ref{s2p24} and \ref{s3p24}, respectively for $\alpha=2$ and $\alpha=3$. In
these cases the approach to the large-$l$ limit is slower, because corrections
are $\Or(l^{-\theta/\alpha})$. As a consequence the $l\to\infty$
extrapolations tend to be more imprecise with increasing $\alpha$, see indeed
the case $\alpha=3$ in figure \ref{s3p24}.  Larger values of $\alpha$ would
require larger trap sizes to get reliable $l\to\infty$
extrapolations, due to the slowly decaying $O(l^{-\theta/\alpha})$ scaling
corrections.  Comparing the results for $\alpha=1,\,2,\,3$, we see that the
functions $f_\alpha(X)$ turn out to depend little on $\alpha$.

\section{The 1D hard-core Bose-Hubbard model with a trapping potential}
\label{sec2}

We now consider the 1D Bose-Hubbard (BH) model (\ref{bhm}), which is of
experimental relevance because it describes quasi-one-dimensional cold atoms
in optical lattices, see, e.g.,
Refs.~\cite{BDZ-08,PWMMFCSHB-04,KWW-04,KWW-05,CFFFI-09}.  We consider its
hard-core limit $U\to\infty$, which implies that the particle number is
restricted to the values $n_i=0,1$, and allows us to study the effects of the
confining potential by very accurate numerical results.

The hard-core limit of the 1D BH model can be mapped into the XX chain model
with a space-dependent transverse external field,
\begin{equation}
H_{\rm XX} = - {J\over 4} \sum_i \left( \sigma^x_i \sigma^x_{i+1} + 
\sigma^y_i \sigma^y_{i+1} \right)
-{1\over 2} \sum_i [\mu+V(x_i)] \sigma^z_i,
\label{XX}
\end{equation}
(which is the Hamiltonian (\ref{Isc}) with $\gamma=0$, apart from a factor 2).
The Pauli spin matrices are related to the boson operators of the BH
Hamiltonian $H_{\rm BH}$, cf.\ (\ref{bhm}), by $\sigma^x_i = b_i^\dagger +
b_i$, $\sigma^y_i = i(b_i^\dagger - b_i)$, $\sigma^z_i = 1-2b_i^\dagger b_i$.
One can then map the XX chain into a model of free spinless fermions by a
Jordan-Wigner transformation, given by (\ref{sfi}) for $\gamma=0$ (apart from
a factor 2).

In the absence of the trap, the 1D hard-core BH model has three phases: two
Mott insulator phases, for $\mu>1$ with $\langle n_i\rangle=0$ (empty state)
and for $\mu<-1$ with $\langle n_i\rangle=1$, separated by a gapless
superfluid phase for $|\mu|<1$ characterized by the filling factor
\begin{equation}
f\equiv \langle n_i\rangle = (1/\pi){\rm arccos} \mu.  
\label{fmu}
\end{equation}
A general analysis of the trap-size dependence and TSS at the Mott transitions
and within the superfluid phase of the hard-core BH model (\ref{bhm}) has been
presented in Ref.~\cite{CV-10-2}.  Here we focus on the behaviour of the
quantum entanglement within the superfluid phase, i.e., for $|\mu|<1$, whose
continuum limit is described by a free massless bosonic field theory with
dynamic exponent $z=1$, thus a conformal field theory with $c=1$.
Specifically, we consider the model at $\mu=0$, corresponding to half filling
in the absence of the trap.

\subsection{Modulated multi-TSS in the superfluid phase}
\label{mmtss}
  
The trap-size dependence shows subtle effects in the parameter region where
the homogeneous model without trap has a nonzero filling $f$, i.e., for
$\mu<1$, including the superfluid phase.  This is essentially related to the
presence of level crossings of the lowest states at finite trap size.  They
arise because the particle number is conserved, i.e., the particle number
operator $\hat{N}=\sum_i n_i$ commutes with the BH Hamiltonian (\ref{bhm})
even in the presence of the trapping potential; thus the eigenvectors do not
depend on $\mu$, even though the eigenvalues do.  In the presence of the
trapping potential (\ref{potential}), the particle number $N\equiv\langle\hat
N\rangle$ is finite and increases as $N\sim l$ with increasing the trap size
$l$, see (\ref{nlrel}).  Therefore, as $l\to\infty$, there is an infinite
number of ground-state level crossings where $N$ jumps by 1 and the gap
vanishes.  In spite of the presence of these level crossings, the length scale
of the critical modes diverges only the in the large trap-size limit.  Note
that the hard-core limit, $U\to \infty$ in (\ref{bhm}), or the spatial
dimension do not play any special role, so we expect that level crossings at
finite trap size are a general feature of the BH model (\ref{bhm}) in the
presence of a confining potential, when the homogeneous limit of infinite trap
size has a finite particle density.  The main effect of the infinite level
crossings in the limit $l\to\infty$ is that the asymptotic power law
behaviours get modulated by periodic functions of the trap size $l$, giving
rise to a modulated TSS.

Moreover, in the superfluid region, the scaling of the spatial dependence of
the correlation functions is characterized by two independent length scales
having different power-law divergences in the large trap-size limit.  One of
them scales as 
\begin{equation}
\xi_s\sim l
\label{sls}
\end{equation}
and describes the behaviour of observables related
to smooth modes, such as the gap and the half-lattice entanglement entropy;
the other one scales as 
\begin{equation}
\xi_F\sim l^\zeta,\qquad \zeta=p/(p+1),
\label{fls}
\end{equation}
and it is
found in observables involving the modes at the Fermi scale $k_F=\pi f$, such
as the density-density correlation.  This gives rise to a multi-TSS
behaviour, which is more complicated than that described by the scaling Ansatz
(\ref{freee}) which applies to the XY chain model around its quantum critical
point.  

As we shall see, the behaviour of the quantum entanglement of parts of
the system in its ground state reflects the above general features.

In the following we briefly mention a number of results obtained in
Ref.~\cite{CV-10-2} which will be useful for the discussion of the quantum
entanglement.

The particle density of the 1D hard-core BH model approaches its local density
approximation (LDA) in the large-$l$ limit, i.e., the value of the particle
density of the homogeneous system at the effective chemical potential
\begin{equation}
\mu_{\rm eff}(x) \equiv \mu +(x/l)^p.
\label{mueff}
\end{equation}
The LDA of the particle density reads
\begin{equation}
\langle n_x \rangle_{\rm lda} \equiv \rho_{\rm lda}(x/l) = 
\kern-10pt \quad\left\{
\begin{array}{l@{\ \ }l@{\ \ }l}
0 & {\rm for} & \mu_{\rm eff}(x) > 1, \\
(1/\pi)\arccos\mu_{\rm eff}(x) &
    {\rm for} & -1 \le \mu_{\rm eff}(x) \le 1, \\
1 & {\rm for} & \mu_{\rm eff}(x) < -1. \\
\end{array} \right.
\label{nxlda}
\end{equation}
Corrections are generally suppressed by powers of the trap size, and present a
nontrivial scaling behaviour~\cite{CV-10-2} in term of the scaling variable
$Y\equiv x/l^\zeta$.

Asymptotically, the total particle number is obtained by integrating the local
density approximation of the particle density $\rho_{\rm lda}$:
\begin{equation}
N \equiv \langle \hat{N} \rangle = c_\mu l + \Or(1) ,
\qquad c_\mu = 2  \int_0^\infty \rho_{\rm lda}(y) \,\rmd y.
\label{nlrel}
\end{equation}
At $\mu=0$ we have $c_{\mu=0}\approx 0.76276$ for $p=2$, $c_{\mu=0}\approx
0.859407$ for $p=4$, and $c_{\mu=0}\to 1$ for $p\to\infty$.

In the superfluid region the interval between subsequent level crossings
$l_0^{(k)}$ (where $k$ enumerates them with increasing trap size) tends to
a constant value in the large trap-size limit, given by~\cite{CV-10-2}
\begin{equation}
\lim_{k\to\infty} \left( l_0^{(k+1)}-l_0^{(k)}\right) = 1/c_\mu, 
\label{levcro}
\end{equation}
with $\Or(l^{-2})$ corrections (this formula has been checked with great
accuracy by numerical computations).  In the limit $p\to\infty$ the interval
of level crossings converges toward the corresponding value in the finite-size
behaviour of the homogeneous system with open boundary conditions, which is
$1/(2f)$.

\begin{figure}[tbp]
\begin{center}
\includegraphics*[scale=\graphicscale]{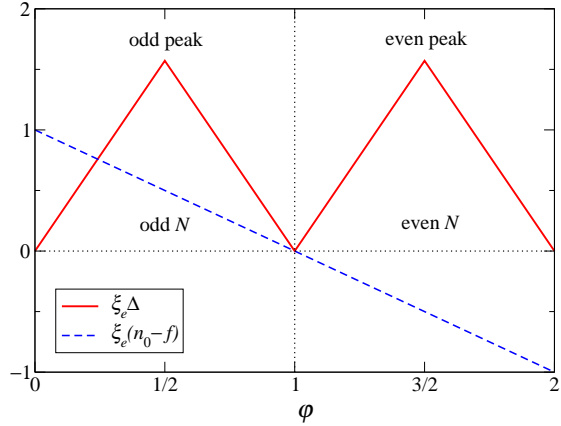}
\end{center}
\vskip-5mm
\caption{
  The asymptotic behaviour of the gap and the particle density at the center
  of the trap at $\mu=0$, thus $f=1/2$. In formulae~\cite{CV-10-2},
  $\Delta=\pi t(\varphi)/\xi_e$ and $\langle n_0 \rangle -f=(1-\varphi)/\xi_e$,
  where $\xi_e$ is the entanglement length scale defined in (\ref{xiemu0}).
  These results apply for any $p$, including $p\to\infty$.  }
\label{gapdens}
\end{figure}

The amplitudes of the asymptotic power law behaviours turn out to be modulated
by periodic functions of the trap size, whose period is related to the level
crossing interval (\ref{levcro}). In order to describe such phenomenon, since
some quantities distinguish between odd and even level crossings, we introduce
the phase-like variable
\begin{equation}
\varphi = 2 {l-l_0^{(2k)}\over  l_0^{(2k+2)} - l_0^{(2k)}}
\qquad {\rm for}\quad l_0^{(2k)}\le l < l_0^{(2k+2)},
\label{barphildef}
\end{equation}
thus $0\le \varphi < 2$. $\varphi$ provides the normalized distance from the
largest even level crossing smaller than the given $l$.\footnote{In
  Ref.~\cite{CV-10-2} the variable $\varphi$ was denoted by $\bar\phi$.}
Values in the range $0\le \varphi \le 1$ corresponds to ground states with odd
particle numbers, while the range $1\le \varphi \le 2$ corresponds to even
particle numbers.  Note that \eref{levcro} implies that $\varphi$ is a
periodic function of $l$ in the asymptotic regime.

In the superfluid phase, i.e. $|\mu|<1$, the asymptotic behaviours of the gap
and the particle density at the center of the trap are given by
\begin{equation}
\eqalign{
\Delta \sim t(\varphi) l^{-1}, \qquad
t(\varphi) = 1/2 - ||1-\varphi|-1/2|, \label{gapXX}\\
\langle n_0 \rangle - f \sim (1-\varphi) l^{-1},}
\label{densXX}
\end{equation} 
respectively.  The amplitudes of the asymptotic power-law behaviours get
modulated by periodic functions of the trap size, through the dependence of
$\varphi$ of their amplitudes. The values $\varphi=1/2$ and $\varphi=3/2$
correspond to the odd and even peaks of the gap, where the particle number
$N=\langle \hat{N} \rangle$ is odd and even respectively.  In
figure \ref{gapdens} we show their behaviours at $\mu=0$.

An analogous modulation of the asymptotic behaviours is found in the limit
$p\to\infty$, i.e. for a homogeneous chain of size $L$ with open boundary
conditions~\cite{CV-10-2}. In this case, the variable $\varphi$ reads
\begin{equation}
\varphi \equiv 2\{[(L+1)f+1]/2\}, 
\label{phipinf}
\end{equation}
where $\{x\}\equiv x - \lfloor x\rfloor$ is the fractional part of $x$ (i.e.,
the sawtooth function). The asymptotic behaviour of the gap is given by
\begin{equation}
\Delta = \pi (1-\mu^2)^{1/2} t(\varphi) L^{-1}.
\label{deltapinf}
\end{equation}
At $\mu=0$, i.e., $f=1/2$, the variable
$\varphi$ takes only discrete values: $\varphi=0,\,1/2,\,1,\,3/2$ (the values
$1/2$ and $3/2$ corresponds to even chain size, with odd and even particle
number respectively).

\subsection{Asymptotic trap-size dependence of the bipartite entanglement 
entropies}
\label{atssRe}

Let us consider a chain with even $L$ sites and open boundary conditions, and
a trap of size $l$ centered between the middle sites of the chain. We again
divide it into two subsystems and study the behaviours of the von Neumann and
R\'enyi entropies of one of the subsystems.  We present results up to trap
sizes $l=\Or(10^3)$, corresponding to total particle numbers of $\Or(10^3)$,
see equation (\ref{nlrel}).  We used chains of size up to $L=3000$ to achieve
negligible finite-$L$ effects (a more detailed discussion of this point for
the XX chain can be found in the App. A of Ref.~\cite{CV-10-2}).
\begin{figure}[tbp]
\begin{center}
\includegraphics*[scale=\graphicscale]{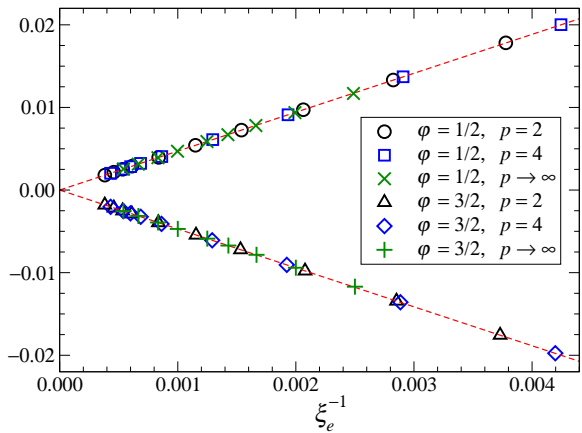}
\end{center}
\vskip-5mm
\caption{The quantity $\xi_e(6 S_{1,1/2} -\ln\xi_e - e_1)$,
  for $\varphi=1/2,\,3/2$ and $p=2,\,4,\,\infty$.  The dashed lines
  show the linear behaviour $\pm 3\pi/(2 \xi_e)$ for
  $\varphi=1\mp1/2$, in agreement with (\ref{s11o2}).  }
\label{s1diff}
\end{figure}

In the absence of the trap, the behaviour of the entanglement entropies is
given by (\ref{ccfo}) with $c=1$, with
\begin{equation}
e_\alpha = \ln\sqrt{1-\mu^2} + \ln (4/\pi) + y_\alpha,
\label{emufunc} 
\end{equation}
obtained using the results of Refs.~\cite{CC-04,JK-04,IJ-08}.  Corrections are again
expected to be $\Or(L^{-1/\alpha})$~\cite{CCEN-10,CC-10}.  The large-$L$
behaviour of numerical results is indeed compatible with the following
asymptotic behaviour
\begin{equation}
\eqalign{\fl S_\alpha(L/2-x;L) = 
C_\alpha \Biggl[ \ln(L+1) +\,\ln\cos (\pi X) + e_\alpha \\
 -
(-1)^{L/2-x} {b_\alpha \over (L \cos\pi X)^{1/\alpha}} + \Or(L^{-2/\alpha})
\Biggr]
}\label{pinfbeh}
\end{equation} 
with $X\equiv x/(L+1)$, and $b_1=3\pi/2\approx4.71239$, $b_2\approx 4.79256$
and $b_3\approx 4.19726$.  Note that the factor $L+1$ in the logarithm and in
the definition of $X$ is essential to get a formula which is correct up to
$\Or(L^{-2})$ at fixed $X$ in the case of the von Neumann entropy ($\alpha =
1$).

\begin{figure}[tbp]
\begin{center}
\includegraphics*[scale=\graphicscale]{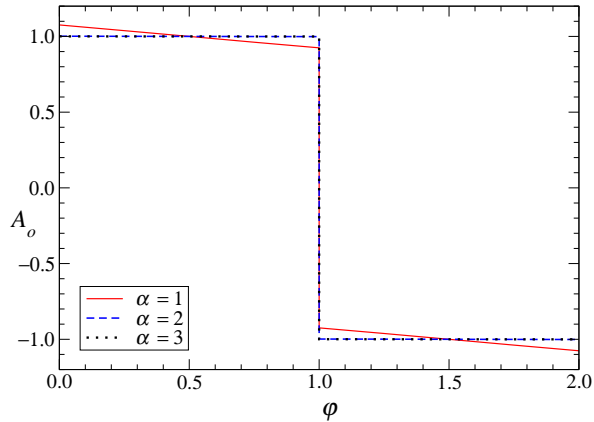}
\end{center}
\vskip-5mm
\caption{The function $A_o(\varphi)$ as obtained from data at $l\approx 998$,
normalized so that $A_o(1/2)=1$ and $A_o(3/2)=-1$.
}
\label{aophi}
\end{figure}

The half-lattice von Neumann entropy in the presence of the trap was already
considered and numerically investigated in Ref.~\cite{CV-10-2}, finding an
entanglement length scale which increases linearly with the trap size in the
whole superfluid region.  Specifically, at $\mu=0$ and for any power $p$ of
the potential, including $p\to\infty$, the half-lattice von Neumann entropy
behaves as~\cite{CV-10-2}
\begin{equation}
S_{1,1/2} = {1\over 6} \left[ \ln\xi_e + e_1 + 
A_o(\varphi){3\pi\over 2\xi_e} +\Or(\xi_e^{-2})\right],
\label{s11o2}
\end{equation}
where
\begin{equation}
\xi_e = a_e l,\qquad 
a_e = \sqrt{\pi}\Gamma\left({1\over 2p}\right)
\left[p \Gamma\left({1+p\over 2p}\right)\right]^{-1}.
\label{xiemu0}
\end{equation}
$\xi_e$
provides an entanglement length scale, and $A_o(\varphi)$ is a periodic
function of $l$ which can be parametrized by
\begin{equation}
A_o(\varphi) = \Biggl\{
\begin{array}{cl}
1 + q (\varphi-1/2) &
\quad {\rm for} \quad 0 < \varphi < 1, \\
-1 - q (3/2-\varphi) &
\quad {\rm for} \quad 1 < \varphi < 2 \\
\end{array}
\label{eq:Ae2l}
\end{equation}
Note that $A_o(1\mp1/2)=\pm1$.
In figure \ref{s1diff} we show results for $p=2,\,4,\,\infty$ and
$\varphi=1/2,\,3/2$.
They clearly support equation \eref{s11o2}.  The function
$A_o(\varphi)$ is shown in figure \ref{aophi} for $p=2$ for trap sizes
around $l\approx 998$.  We estimate 
$q\approx -0.152$ for $p=2$ and $q\approx -0.044$ for $p=4$.  The
parametrization \eref{eq:Ae2l} of $A_o(\varphi)$ is convenient when comparing
with the $p\to\infty$ limit, because $\varphi=1/2,\,3/2$ are the only possible
values for the homogeneous system in a chain with even $L$, corresponding to
odd and even $L/2$ (thus odd and even particle number $N$) respectively.  The
behaviour of the homogeneous system with open boundary condition, cf.\ 
(\ref{ccfo}), is obtained by replacing $\xi_e\to L+1$ at the values
$\varphi=1/2,\,3/2$.

\begin{figure}[tbp]
\begin{center}
\includegraphics*[scale=\graphicscale]{fig11a.eps}
\includegraphics*[scale=\graphicscale]{fig11b.eps}
\end{center}
\vskip-5mm
\caption{The quantity $H_\alpha$, defined in (\ref{Hdef}), for
  $\alpha=2,\,3$, $\varphi=1/2,\,3/2$ and $p=2,\,4,\,\infty$.  The dashed
  lines show the linear behaviour $\pm b_\alpha \xi_e^{-1/\alpha}$ for
  $\varphi=1\mp1/2$, with $b_2=4.79256$ and $b_3=4.19726$, cf.\ (\ref{sa1o2}).
}
\label{salphadiff}
\end{figure}

\begin{figure}[tbp]
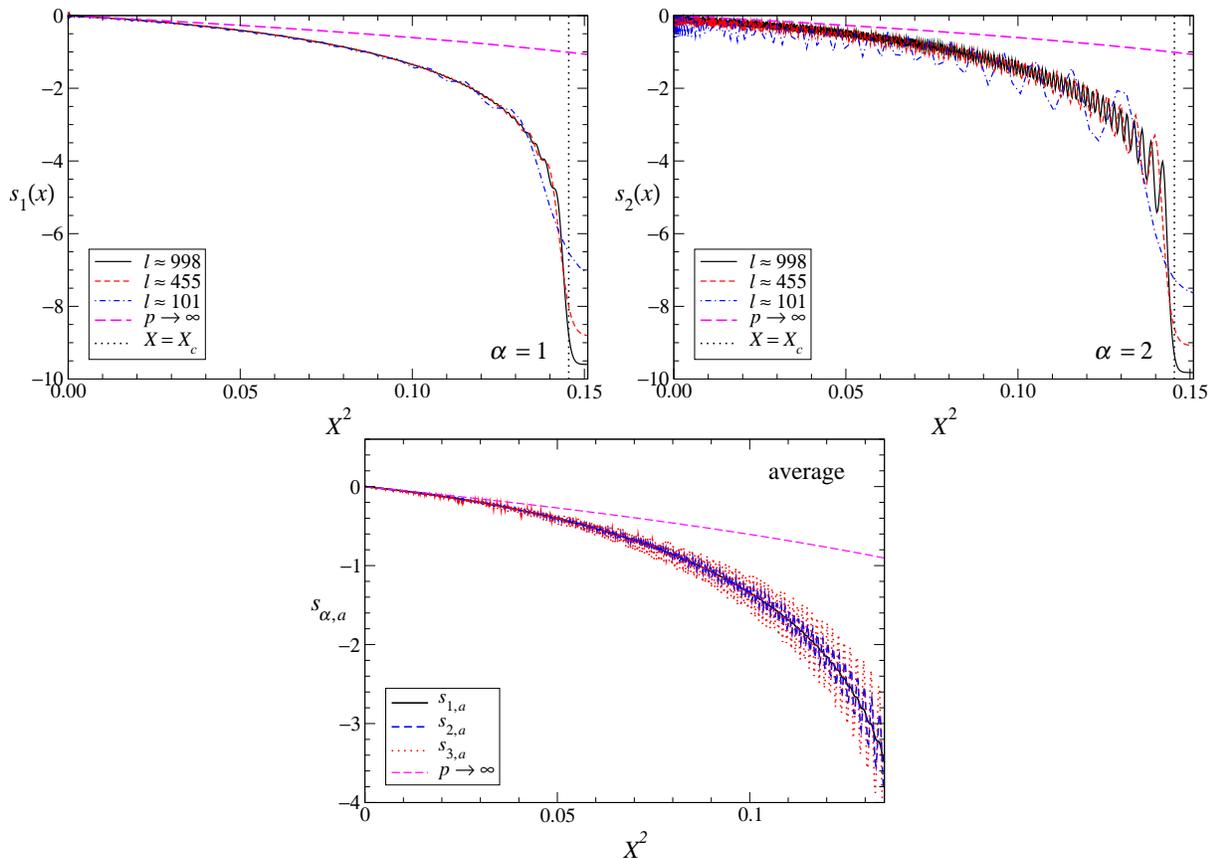

\begin{center}
\hbox{\includegraphics*[scale=\smallgraphicscale]{fig12a.eps}
\includegraphics*[scale=\smallgraphicscale]{fig12b.eps}}
\includegraphics*[scale=\smallgraphicscale]{fig13c.eps}
\end{center}
\vskip-5mm
\caption{$s_\alpha$ vs.\ $X^2$, for $\alpha=1,\,2,\,3$, $p=2$,
   $\varphi=1/2$, and a few values of $l$.  The vertical dotted
   lines show the value $X=X_c=1/a_e$, cf. (\ref{xiemu0}).
}
\label{salphaXX}
\end{figure}

We now show that the half-lattice R\'enyi entropies show analogous behaviours.
In figure \ref{salphadiff} we show numerical results of the quantity
\begin{equation}
H_{\alpha}\equiv \xi_e^{1/\alpha} (C_\alpha^{-1} S_{\alpha,1/2} -\ln\xi_e - e_\alpha)
\label{Hdef}
\end{equation}
for $p=2,\,4,\,\infty$ and $\varphi=1/2,\,3/2$.  The data
appears to follow curves which are apparently independent of $p$, but clearly
depending on $\varphi$.  Therefore, analogously to the case of the von Neumann
entropy, the asymptotic behaviour of the R\'enyi entropies show the same form
for any value of the potential power $p$, including $p\to\infty$, when they
are expressed in terms of the entanglement length scale $\xi_e$, i.e.,
\begin{equation}
S_{\alpha,1/2} = C_\alpha \left[ \ln\xi_e + e_\alpha + 
b_\alpha A_{\alpha,o}(\varphi) \xi_e^{-1/\alpha} +\Or(\xi_e^{-2/\alpha})\right]
\label{sa1o2}
\end{equation}
where $b_\alpha$ are the same constants entering the behaviour of the
half-lattice R\'enyi entanglement entropies of the homogeneous system,
cf.\ (\ref{pinfbeh}), and $A_{\alpha,o}$ is given by (\ref{eq:Ae2l}), with
very small values of $q$, i.e.  $q\approx -0.003$ for $\alpha=2$,
$p=2$ and $q\approx-0.001$ for $\alpha=3$, $p=2$.
Thus $A_{\alpha,o}\approx (-1)^N$ for $\alpha>1$.
Equations (\ref{s11o2}) and (\ref{sa1o2}) imply that the length scales
defined in (\ref{ccfot}) behave as
\begin{equation}
  \xi_\alpha = \xi_e \left[ 1  + \Or(\xi_e^{-1/\alpha})\right] 
\label{xialpha}
\end{equation}
for any $\alpha$.  

\begin{figure}[tbp]
\begin{center}
\hbox{\includegraphics*[scale=\smallgraphicscale]{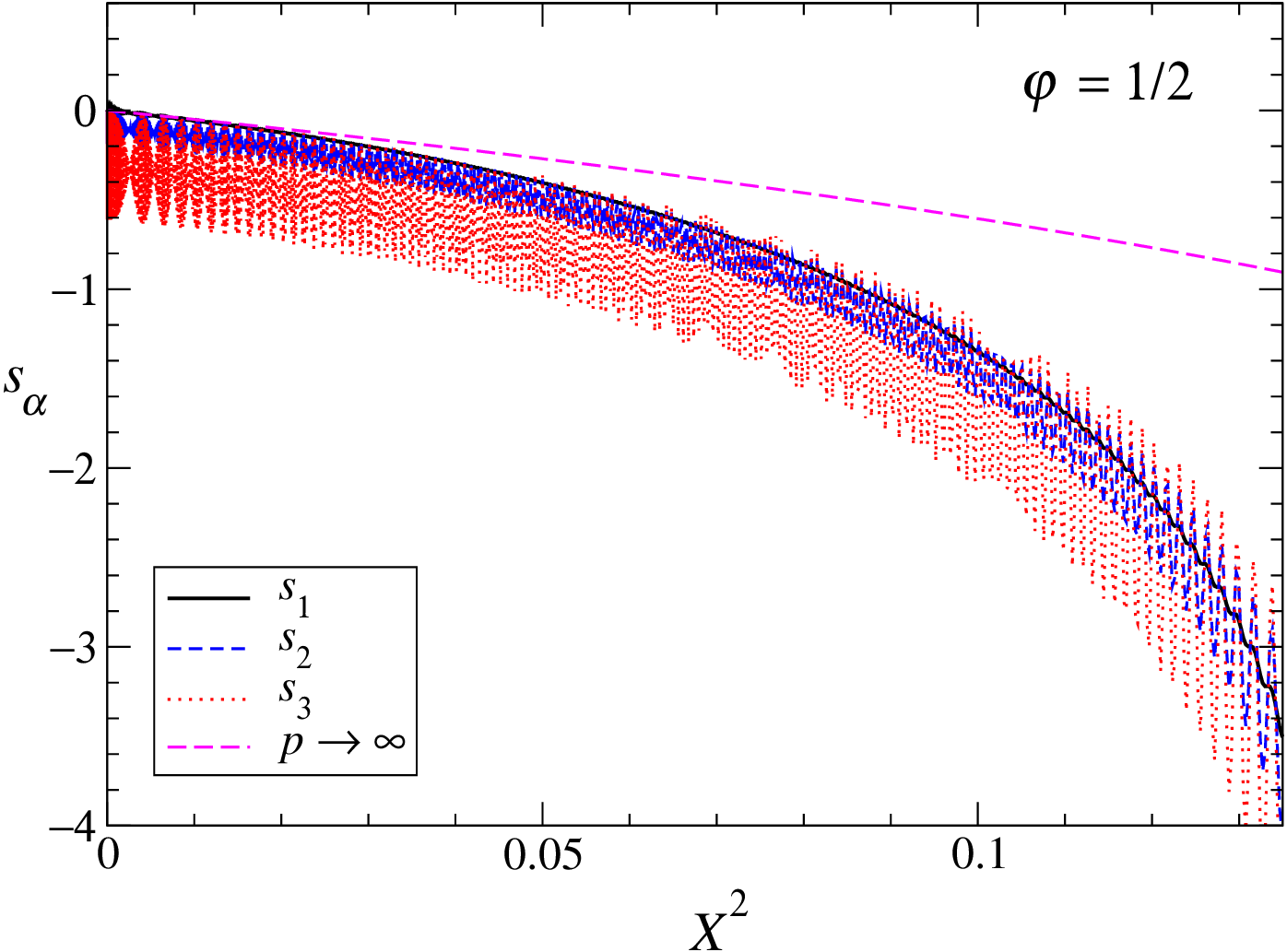}
\includegraphics*[scale=\smallgraphicscale]{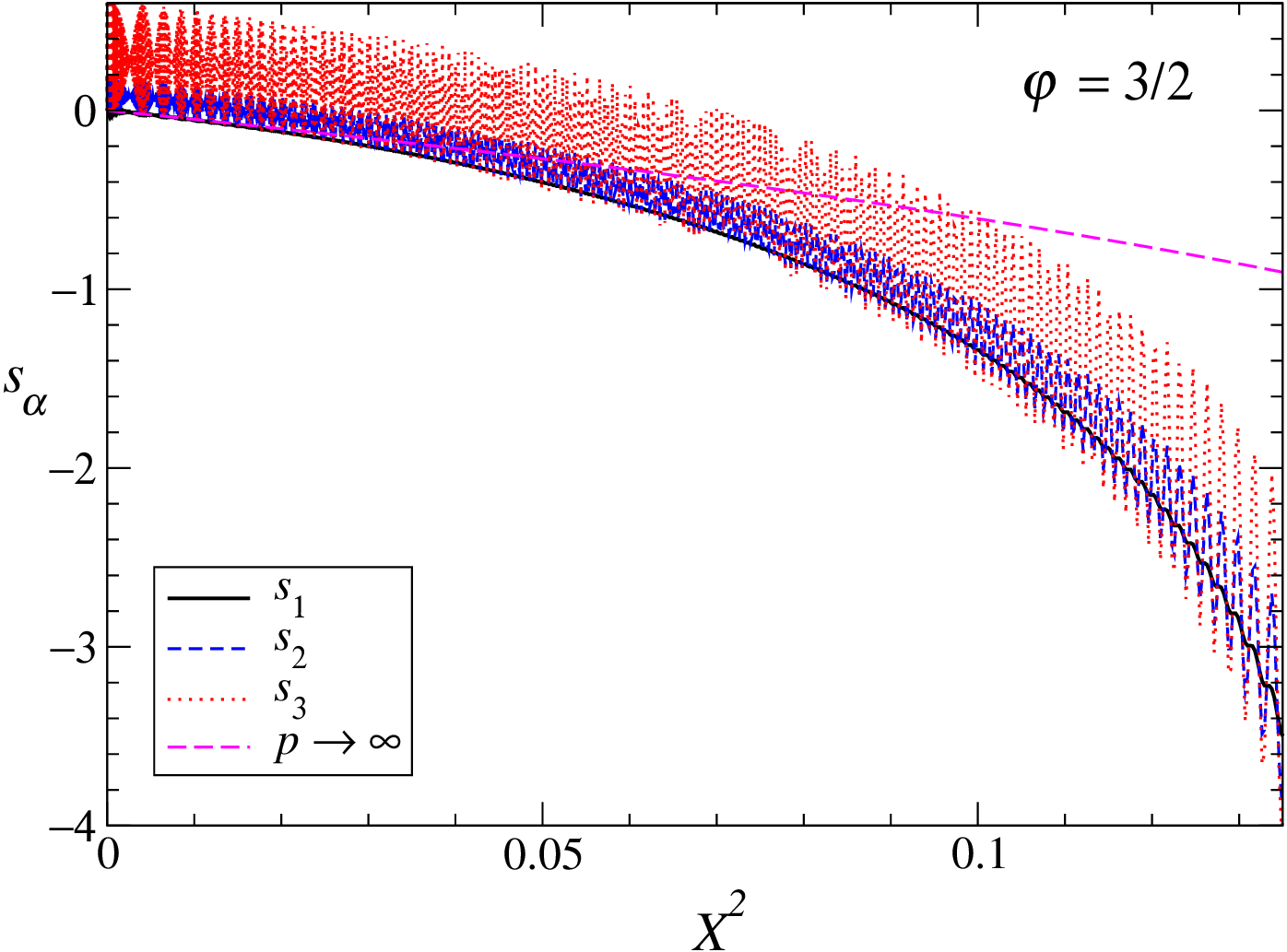}}
\includegraphics*[scale=\smallgraphicscale]{fig13c.eps}
\end{center}
\vskip-5mm
\caption{$s_\alpha$ vs.\ $X^2$ for $\alpha=1,2,3$,
$p=2$, and $l\approx 10^3$, at $\varphi=1/2$, $\varphi=3/2$,
corresponding to $N=761,\,762$, and their average
$s_{\alpha,a}\equiv(s_\alpha|_{\varphi=1/2}+s_\alpha|_{\varphi=1/3})/2$,
which has much smaller oscillations.}
\label{s123XX}
\end{figure}

\begin{figure}[tbp]
\begin{center}
\hbox{\includegraphics*[scale=\graphicscale]{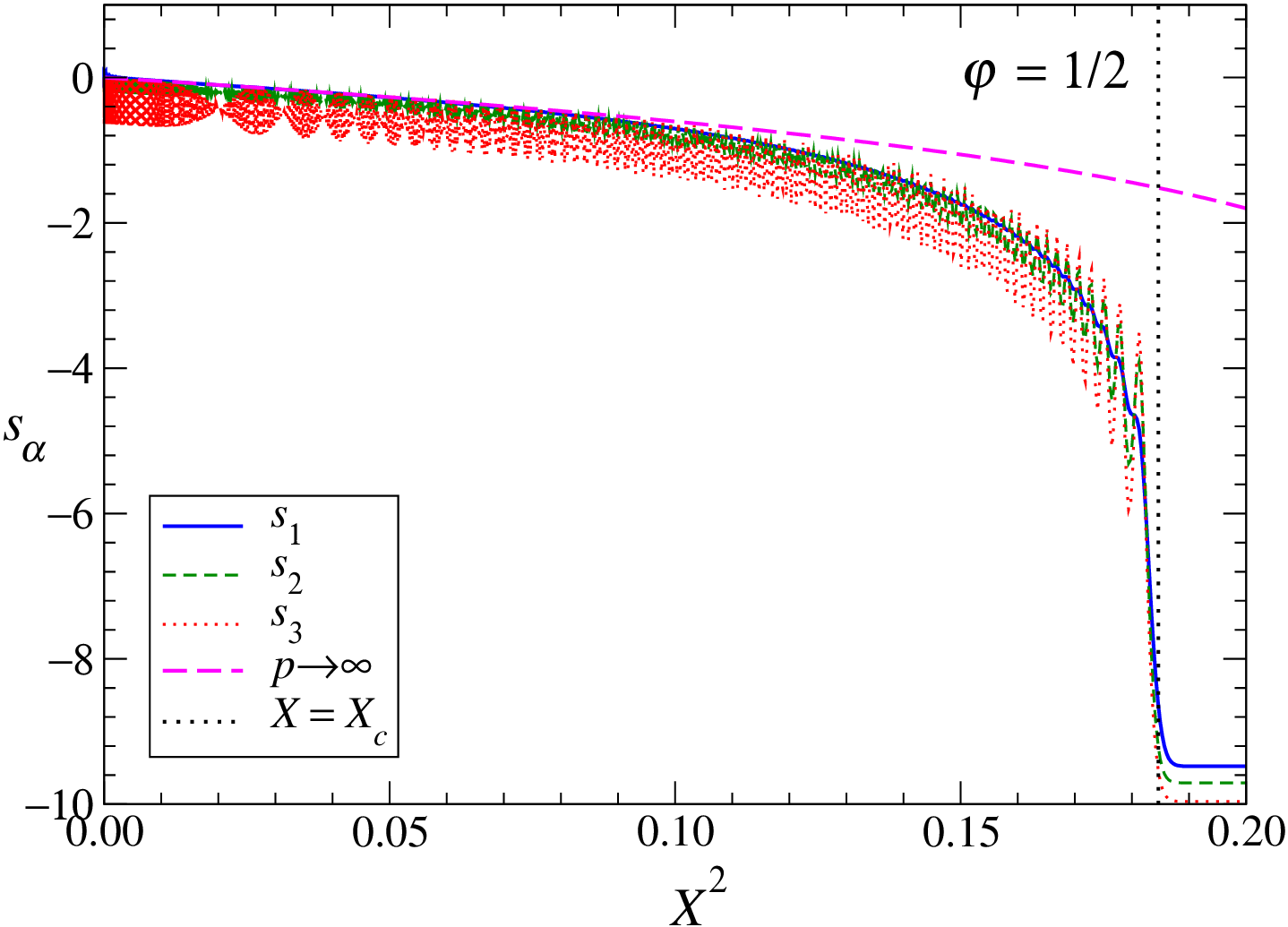}
\includegraphics*[scale=\graphicscale]{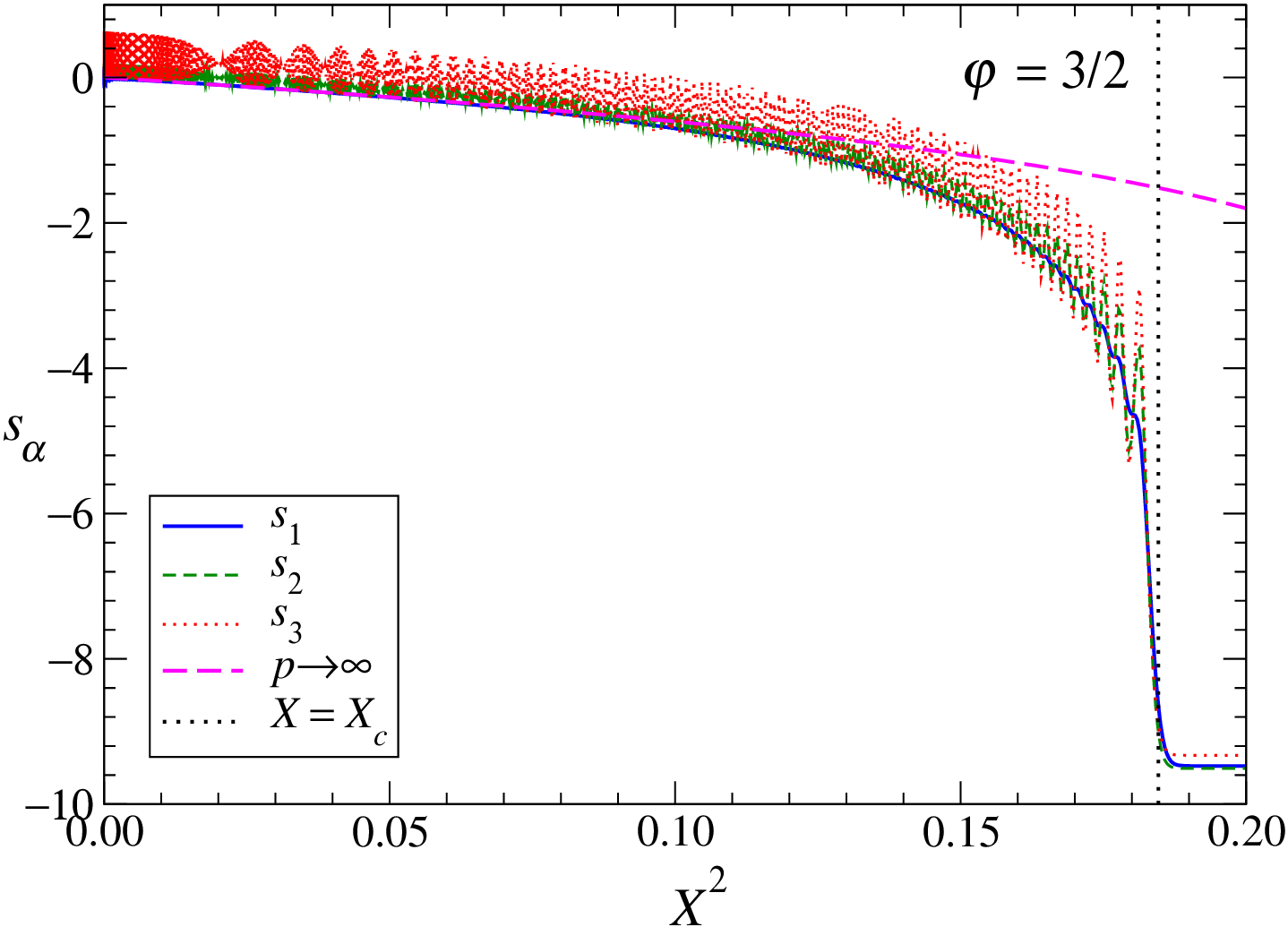}}
\end{center}
\vskip-5mm
\caption{$s_\alpha$ vs.\ $X^2$ for $\alpha=1,2,3$, $p=4$,
   $l\approx 10^3$, and $\varphi=1/2,\,3/2$,
   corresponding to $N=857,\,856$.
The vertical dotted
   lines show the value $X=X_c=1/a_e$, cf. (\ref{xiemu0}).
}
\label{s123XXp4}
\end{figure}

We now analyze the spatial dependence of the bipartite entanglement entropies.
We again consider the functions defined in (\ref{xdepstrap}).  In
figures \ref{salphaXX}, \ref{s123XX}, and \ref{s123XXp4} we show results for
$p=2,\,4$, $\alpha=1,\,2,\,3$, $\varphi=1/2,\,3/2$, and 
several values of the trap size. 
In all cases the behaviours of the functions $s_\alpha(x)$, defined as in
(\ref{xdepstrap}), are characterized by a smooth function $f_\alpha(X)$
with $X\equiv x/\xi_e$, and oscillations around it, which appear
larger and larger with increasing $\alpha$, but get suppressed with increasing
$\xi_e$.  These numerical results suggest
\begin{equation}
s_\alpha(x)=f_\alpha(X)+ \Or(\xi_e^{-1/\alpha}),\qquad X\equiv x/\xi_e,
\label{salpha1}
\end{equation}
where the power-law of the corrections can be guessed from the fact that
$\Or(L^{-1/\alpha})$ corrections are expected in homogeneous system with open
boundary conditions~\cite{CCEN-10,CC-10}.  Moreover, the numerical results
strongly suggest that a unique function describes the leading behaviour for any
$\alpha$, i.e., $f_\alpha(X)\equiv f(X)$ is independent of $\alpha$, for any
$p$.  This becomes more evident when considering the average $s_{\alpha,a}$
of the functions
$s_\alpha(x)$ at subsequent odd and even peaks of the gap (indeed, as we shall
see below, the curves at $\varphi=1/2$ and $\varphi=3/2$ have opposite
oscillating corrections at least for sufficiently small values of $X$).  With
increasing $p$ the function $f(X)$ appears to tend to the $p\to\infty$ limit
given by (\ref{largeplimit}).  Actually, the small-$X$ behaviour at $p=2$
is already very close to the $p\to\infty$ limit (the large-$l$ extrapolation
of the coefficient $c_2$ of the $\Or(X^2)$ term is compatible with its
$p\to\infty$ limit $c_2=-\pi^2/2$ already at $p=2,\,4$).  On the other hand,
oscillations are strongly dependent on $\alpha$.

\begin{figure}[tbp]
\begin{center}
\includegraphics*[scale=\graphicscale]{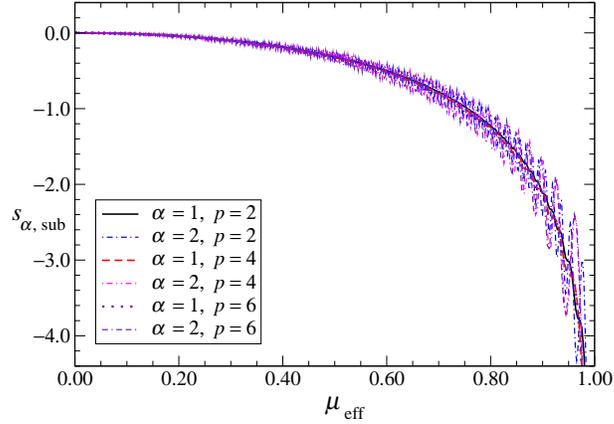}
\end{center}
\vskip-5mm
\caption{$s_{\alpha,sub}\equiv {1\over2}
  [s_\alpha(x)|_{\varphi=1/2} +s_\alpha(x)|_{\varphi=3/2}]-\ln\cos\pi X$
  vs.\ $\mu_{\rm eff}=(x/l)^p$ for $\alpha=1,2$
  and $p=2,\,4$ (for $l\approx 1000$) and $p=6$ (for $l\approx 1400$).}
\label{sasub}
\end{figure}

An interesting numerical relation concerning the leading behaviour $f(X)$ is
illustrated in figure \ref{sasub}: for $p=2,\,4,\,6$, we find that the
difference between $f(X)$ and its $p\to\infty$ limit given by CFT, i.e.
$\lim_{p\to\infty} f(X)=\ln\cos\pi X$, is a function of $\mu_{\rm eff}\equiv
(|x|/l)^p=(a_e X)^p$ only, i.e.,
\begin{equation}
f(X) \approx \ln\cos\pi X + A(\mu_{\rm eff}) 
\label{amueff}
\end{equation}
where the ($p$-independent) function $A$ behaves as $A(\mu_{\rm
  eff})=\Or(\mu_{\rm eff}^2)$ at small $X$ (more precisely the numerical
results indicate $A(\mu_{\rm eff}) = a \mu_{\rm eff}^2+O(\mu_{\rm eff}^4)$ with
$a\approx -1.0$).

\begin{figure}[tbp]
\begin{center}
\includegraphics*[scale=\graphicscale]{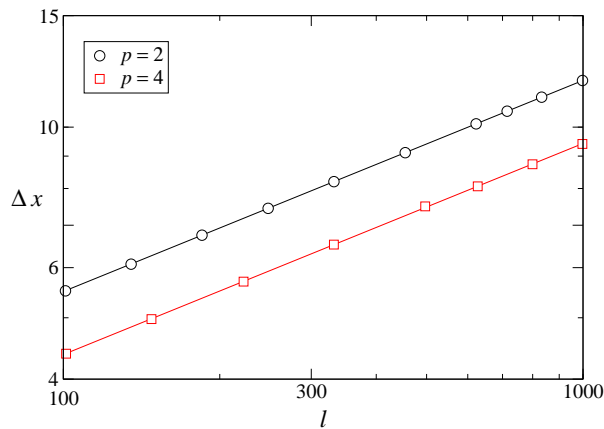}
\end{center}
\vskip-5mm
\caption{
  $\Delta x$ vs.\ $l$ for $\alpha=2$, $p=2,\,4$ and $\varphi=1/2$.  The
  solid lines are one-parameter fits with the form $a l^{1/3}$, cf.\ 
  (\ref{enleq}).
}
\label{s2Delta}
\end{figure}

Note that the von Neumann and R\'enyi entropies rapidly decrease around
$X=X_c$ corresponding to $x_c\equiv l$, thus $X_c=1/a_e$, cf.\ (\ref{xiemu0});
$x_c$ is the value of $x$ where $\mu_{\rm eff}(x) = (x/l)^p = 1$, i.e., the
value of the chemical potential corresponding to the superfluid to empty state
transition, where the particle density of the ground state vanishes, see
equation (\ref{nxlda}).  We thus expect that, for generic values of $\mu$ and
$p$, the region around $x=x_c$ develops critical modes related to a
low-density Mott transition from the superfluid phase to the empty state.  The
effective chemical potential can be expanded around $x_c$ as
\begin{equation}
\mu_{\rm eff}=(x/l)^p 
= 1 + p{x-x_c\over l} + O[(x-x_c)^2].
\label{linV}
\end{equation}
Thus, the behaviour around $x_c$ is essentially analogous to that arising at
the low-density Mott transition $\mu=1$ in the presence of a linear potential
$V_l\sim r/l$.  Around $x_c$, critical modes should appear with length scale
$\xi\sim l^\sigma$, where $\sigma$ is the exponent associated with a linear
external potential.  The value of $\sigma$ can be inferred by RG arguments
analogous to those leading to the determination of the trap exponent $\theta$ at
the low-density Mott transition~\cite{CV-10}, which give
$\sigma=1/3$.\footnote{The exponent $\sigma$ can be determined by a RG
  analysis of the perturbation corresponding to a linear potential
  $V_l(x)=ux$, i.e., $\int \rmd^dx\,\rmd t\, V_l(x) |\phi(x,t)|^2$, at the
  fixed point 
  of the continuous theory describing the Mott transition~\cite{FWGF-89}.  The
  exponent $\sigma$ is related to the RG dimension $y_u$ of the parameter $u$,
  which can be obtained from the relations $y_u - 1 = d+z-y_{|\phi|^2} =
  y_\mu=2$, thus $y_u=3$, and therefore $\sigma\equiv 1/y_u=1/3$ for $d=1$ and
  $d=2$.}  We thus expect that the transition region around $x=x_c$ enlarges
as
\begin{equation}
\Delta x \sim l^{1/3}, 
\label{enleq}
\end{equation}
independently of the power-law $p$ of the confining potential.  This
behaviour is confirmed by the numerical values of
$\Delta x\equiv l-x_{\rm max}$, where
$x_{\rm max}$ is the abscissa of the rightmost maximum of $s_\alpha(x)$,
interpolated to continuous $x$; figure \ref{s2Delta} shows the perfect 
agreement with (\ref{enleq}).

\begin{figure}[tbp]
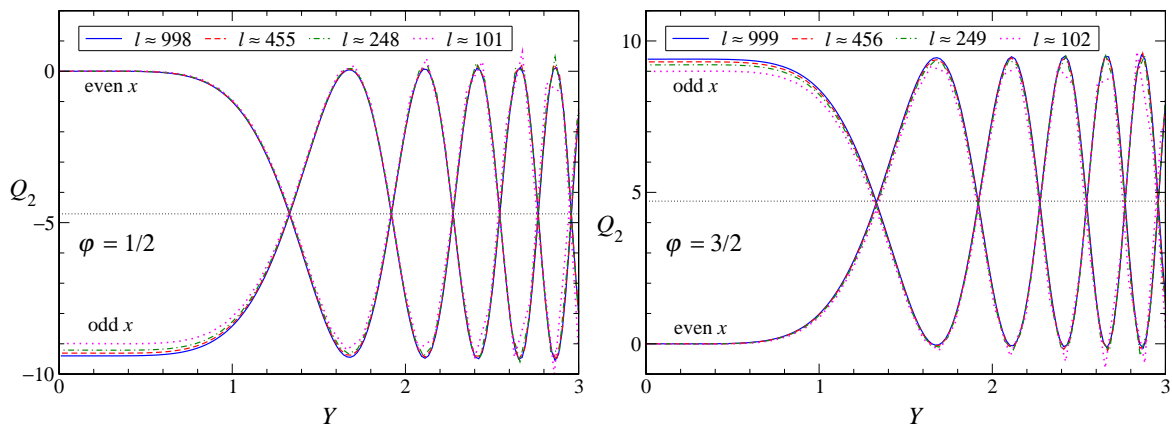

\begin{center}
\includegraphics*[scale=\graphicscale]{fig17a.eps}
\includegraphics*[scale=\graphicscale]{fig17b.eps}
\end{center}
\vskip-5mm
\caption{
  The function $Q_2(x)$, cf.\ (\ref{ralphadef}), vs.\ 
  $Y\equiv x l^{-2/3}$, for $p=2$ and $\varphi=1/2,\,3/2$.
}
\label{s2lms1}
\end{figure}

\begin{figure}[tbp]
\begin{center}
\includegraphics*[scale=\graphicscale]{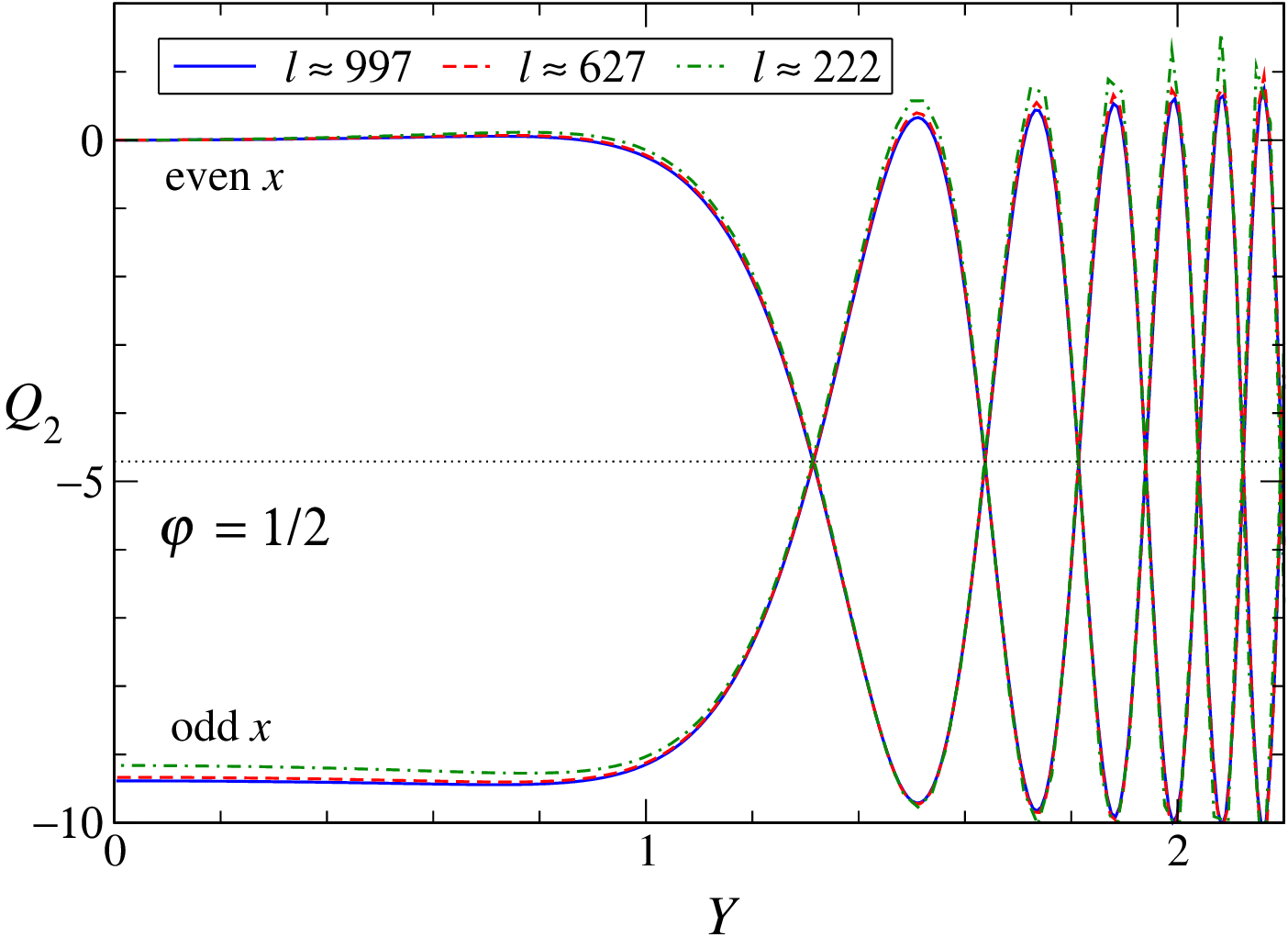}
\includegraphics*[scale=\graphicscale]{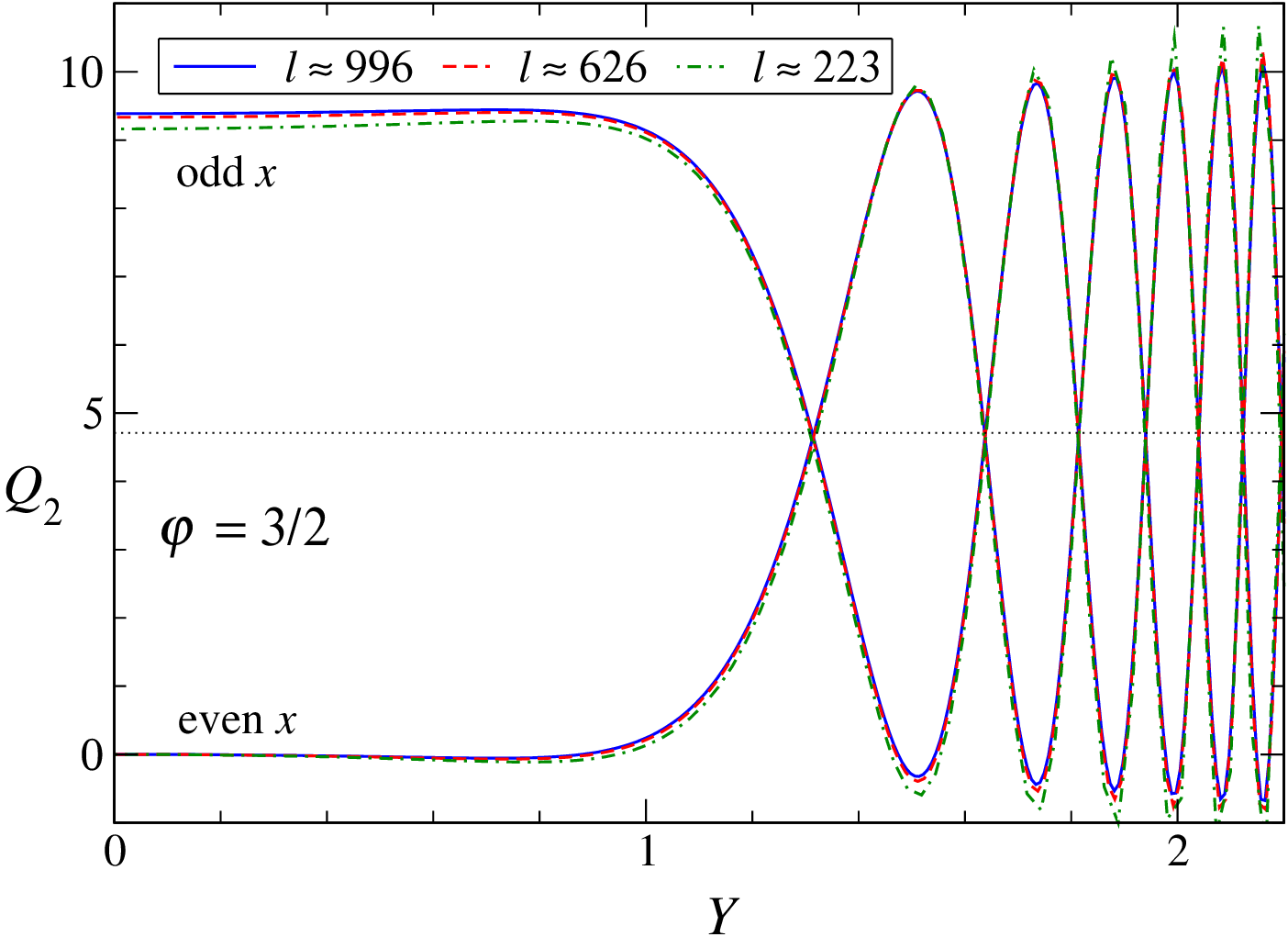}
\end{center}
\vskip-5mm
\caption{
  The function $Q_2(x)$, cf.\ (\ref{ralphadef}), vs.\ 
  $Y\equiv x l^{-4/5}$, for $p=4$ and $\varphi=1/2,\,3/2$. 
}
\label{s2lms1p4}
\end{figure}

\begin{figure}[tbp]
\begin{center}
\includegraphics*[scale=\graphicscale]{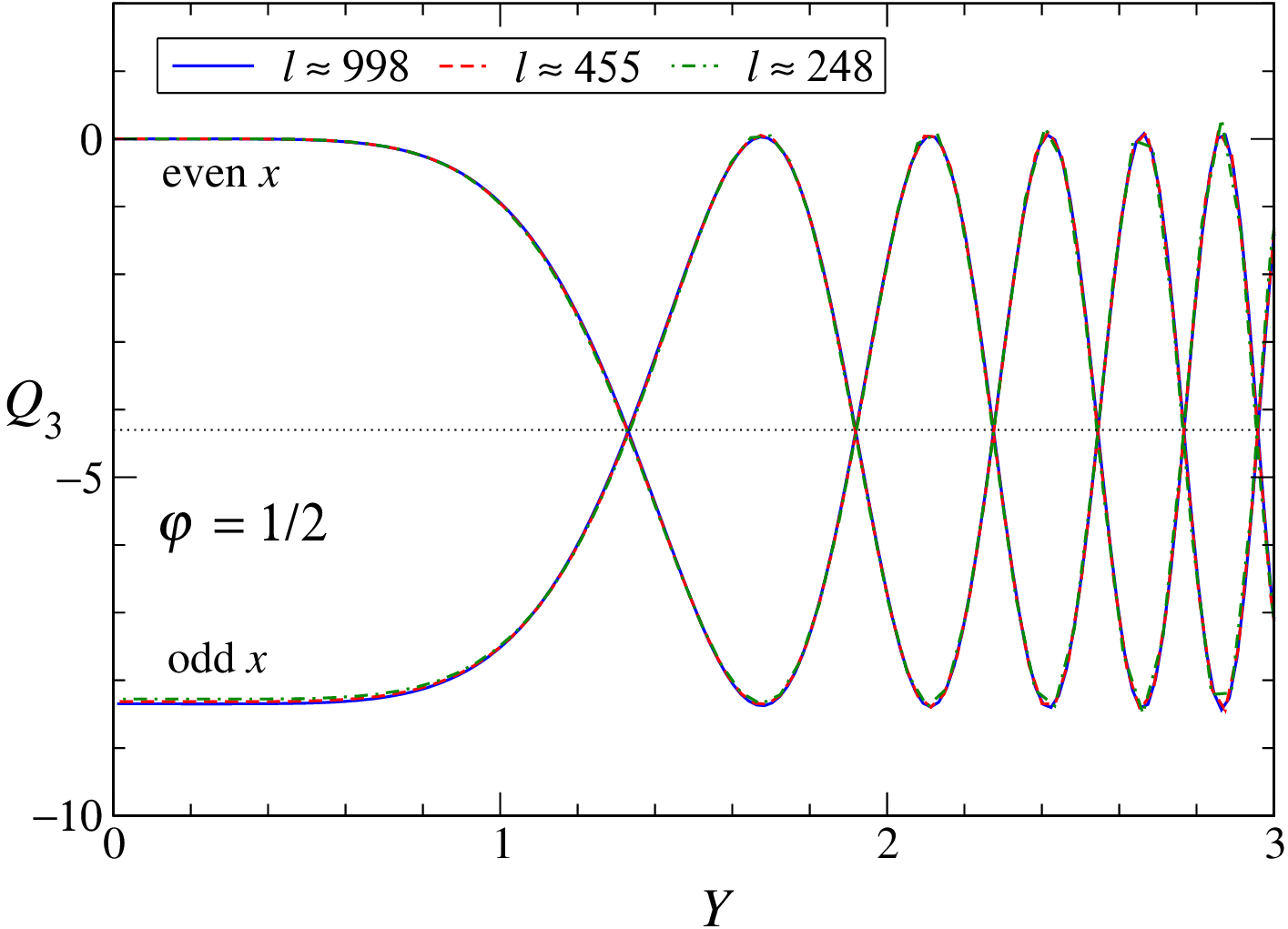}
\includegraphics*[scale=\graphicscale]{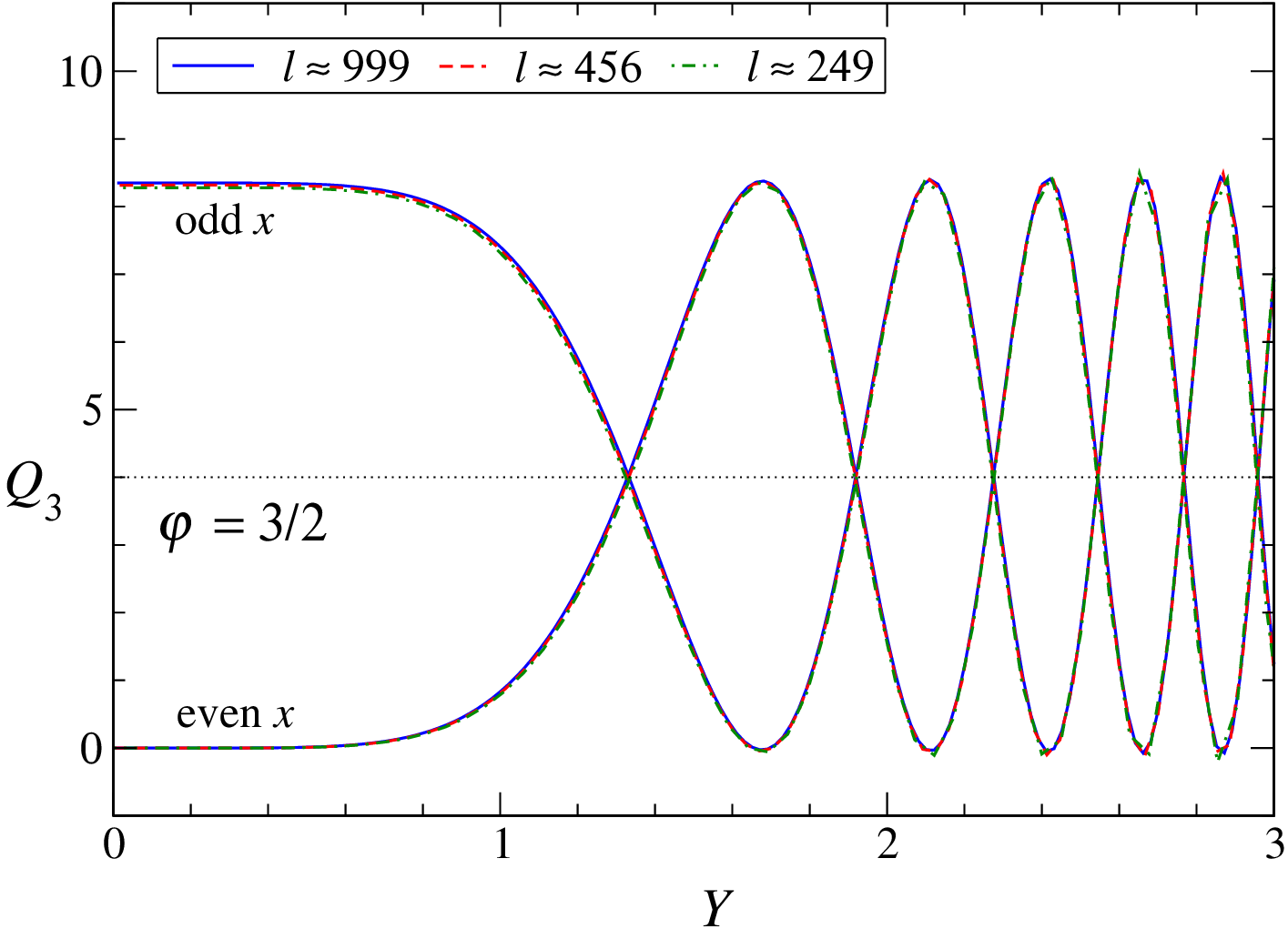}
\end{center}
\vskip-5mm
\caption{
  The function $Q_3(x)$, cf.\ (\ref{ralphadef}), vs.\ $Y\equiv
  x l^{-2/3}$, for $p=2$ and $\varphi=1/2,\,3/2$.
}
\label{s3lms1}
\end{figure}

The behaviours of the subleading and oscillating $\Or(\xi^{-1/\alpha})$
corrections are instead more subtle.  In order to study them, we consider the
function
\begin{equation}
  Q_\alpha(x) = \xi_e^{1/\alpha} \left[ s_\alpha(x) - s_1(x)\right] 
\label{ralphadef}
\end{equation}
for $\alpha>1$. Assuming that the leading behaviour is given by the same
function $f(X)$ for any $\alpha$, $Q_\alpha$ should provide the
$\Or(\xi_e^{-1/\alpha})$ term in (\ref{salpha1}).  The functions
$Q_\alpha(x)$ for $\alpha=2,\,3$, $p=2,\,4$, and $\varphi=1/2,\,3/2$
are shown in figures \ref{s2lms1}, \ref{s2lms1p4} and \ref{s3lms1}, as
obtained by numerical results at fixed trap size for 
$10^2\lesssim l\lesssim 10^3$. 
They clearly show the expected even-odd spatial effect proportional to
$(-1)^x=\rme^{2\rmi k_Fx}$, but also a $\varphi$-dependent scaling limit as a
function of $Y=x/l^\zeta$ with $\zeta=p/(p+1)$.  More precisely, the numerical
results show that the asymptotic behaviour of the the functions $Q_\alpha(x)$
can be written as
\begin{equation}
Q_\alpha(x) \approx (-1)^x g_{\alpha,o}(Y,\varphi) +  g_{\alpha,s}(Y,\varphi)
\label{ralphabeh}
\end{equation}
apart from further suppressed scaling corrections. Note that this strong
evidence of scaling provides also an indirect and accurate check of the
initial hypothesis that the leading behaviour as a function of $X$ does not
depend on $\alpha$, as in the $p\to\infty$ limit.  Summarizing, we have
\begin{equation}
s_\alpha(x) = f(X) + \xi_e^{-1/\alpha} 
\left[ (-1)^x g_{\alpha,o}(Y,\varphi) + 
g_{\alpha,s}(Y,\varphi)\right] + \Or(\xi_e^{-2/\alpha})
\label{genbeh}
\end{equation}
where $X\equiv x/\xi_e$ and $Y\equiv x/l^\zeta$.  We note that the behaviour
of the von Neumann entropy, $\alpha\to 1$, is quite similar to that of the
particle density, which turns out to be~\cite{CV-10-2}
\begin{equation}
\langle n_x\rangle \approx 
\rho_{\rm lda}(X) + \xi_e^{-1} \left[ (-1)^x h_o(Y,\varphi) +
h_s(Y,\varphi)\right] \label{eq:rhox,sf}
\end{equation}

\begin{figure}[tbp]
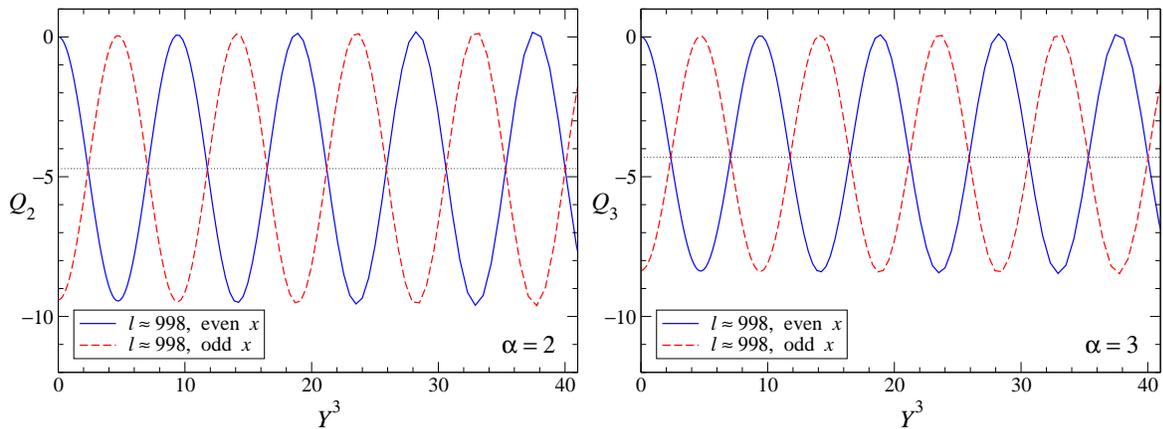

\begin{center}
\includegraphics*[scale=\graphicscale]{fig20a.eps}
\includegraphics*[scale=\graphicscale]{fig20b.eps}
\end{center}
\vskip-5mm
\caption{
The functions $Q_\alpha(x)$, cf.\ (\ref{ralphadef}), vs.\ 
$Y^3 \equiv x^3/l^2$, for $p=2$, $\varphi=1/2$, and $\alpha=2,\,3$.
}
\label{s23ms1y3}
\end{figure}

\begin{figure}[tbp]
\begin{center}
\includegraphics*[scale=\graphicscale]{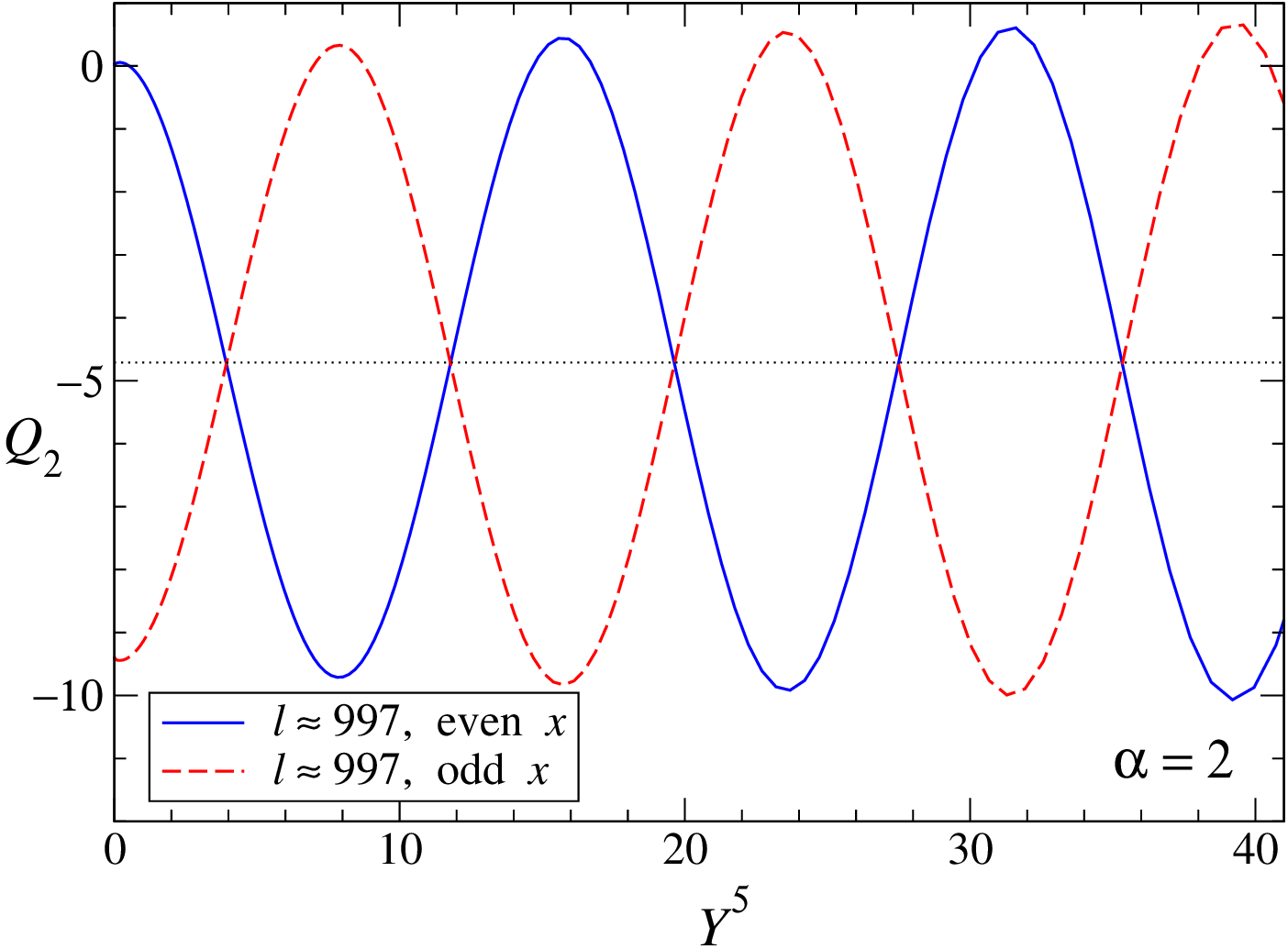}
\includegraphics*[scale=\graphicscale]{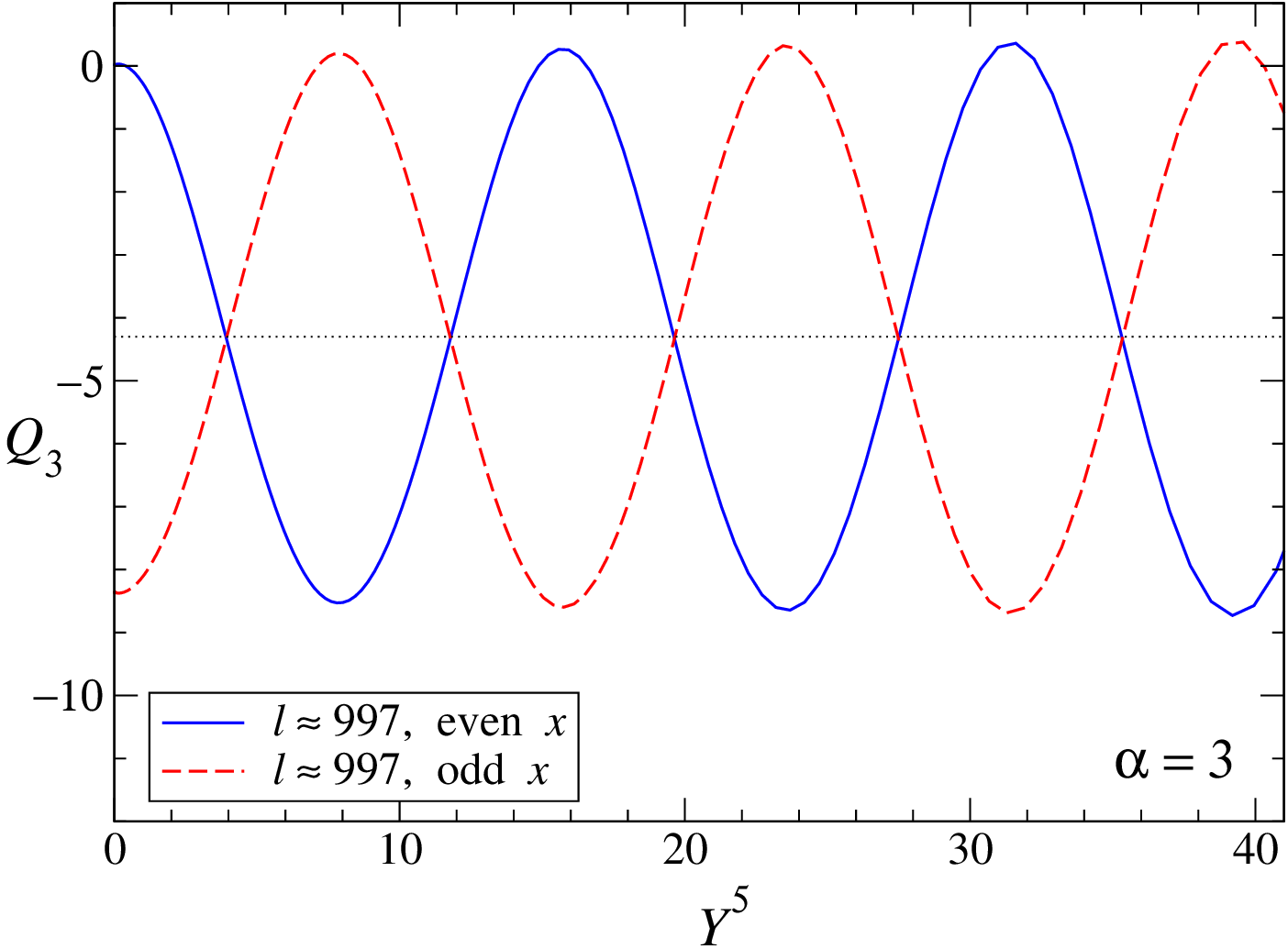}
\end{center}
\vskip-5mm
\caption{
  The functions $Q_\alpha(x)$, cf.\ (\ref{ralphadef}), vs.\ $Y^5 \equiv
  x^5/l^4$, for $p=4$, $\varphi=1/2$, and $\alpha=2,\,3$.
}
\label{s23ms1p4y5}
\end{figure}

\begin{figure}[tbp]
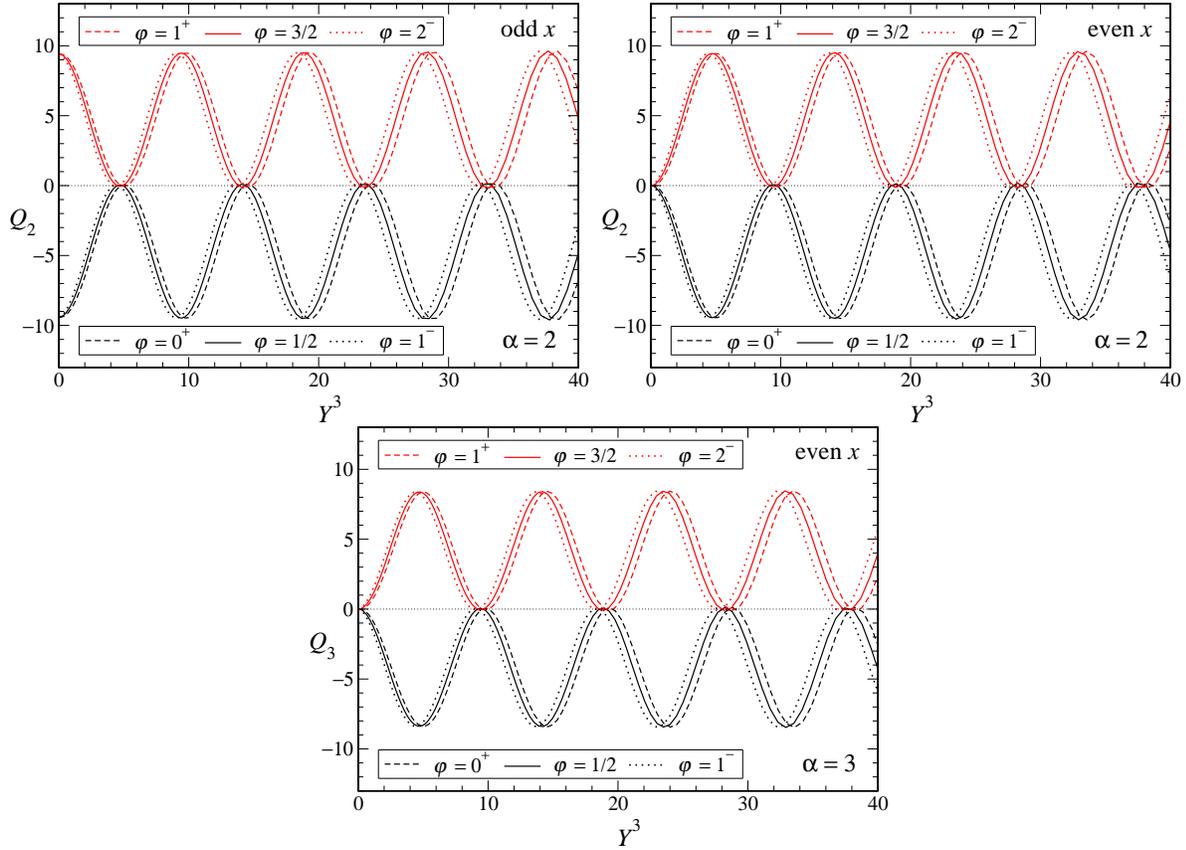

\begin{center}
\hbox{\includegraphics*[scale=\graphicscale]{fig22a.eps}
\includegraphics*[scale=\graphicscale]{fig22b.eps}}
\includegraphics*[scale=\graphicscale]{fig22c.eps}
\end{center}
\vskip-5mm
\caption{The functions $Q_\alpha(x)$, cf.\ (\ref{ralphadef}),
  vs.\ $Y^3 \equiv x^3/l^2$, for $p=2$,
  $\alpha=2,\,3$, several values of $\varphi$, for sites with
  even and odd parity, and trap sizes $l\approx 998$.
  $Q_\alpha$ as a function of $\varphi$ has jumps for integer
  $\varphi$; we denote $\lim_{\varphi{\scriptscriptstyle\searrow}n}$ 
  by $\varphi=n^+$
  and $\lim_{\varphi{\scriptscriptstyle\nearrow}n}$ by $\varphi=n^-$.
  }
\label{s2ms1phi}
\end{figure}

Moreover, as shown by figures \ref{s23ms1y3} and \ref{s23ms1p4y5}, for
$p=2$ and $p=4$ respectively, the oscillations appear to be very close
to periodic functions when plotted versus $Y^{p+1}$.
This is particularly verified in the case of the harmonic potential,
where the data at the odd and even peaks of the gap, corresponding to
ground states with odd and even particle number $N$, accurately fit
the simple function
\begin{equation}
Q_\alpha(x)|_{\varphi=1/2,\,3/2} \approx 
(-1)^N b_\alpha
\left[(-1)^x \cos(c Y^3) -1\right] 
\label{rap2}
\end{equation}
where $b_\alpha$ are the same constants entering the behaviour of the R\'enyi
entropies of the homogeneous system, and $c\approx2/3$.  In
figure \ref{s2ms1phi} we show data for other values of $\varphi$.  Apart from
the prefactor $(-1)^N$,
there is a small residual dependence on
$\varphi$, which is apparently relegated to a small dependence of
$c$ on $\varphi$.

\section{Conclusions}
\label{conclusions}

In this paper we have investigated the scaling behaviour of the bipartite
entanglement of 1D lattice systems in the presence of a power-law confining
potential $V(r) = (r/l)^p$, at a quantum critical point.  We have considered
lattice systems whose quantum critical behaviours are described by 2D conformal
field theories in the absence of the trap, thus showing logarithmically
divergent entanglement entropies, see, e.g., (\ref{ccfo}).

We have considered 1D lattice models of even size $L$ and open boundary
conditions, with a trap of size $l$ centered between the middle sites of the
chain.  We have divided the chain in two parts of length $l_A\equiv L/2-x$ and
$L-l_A$, and computed their von Neumann and R\'enyi entropies
$S_\alpha(L/2-x;L)$.  In the presence of a trap of size $l$, the dependence on
$L$ disappears when $L\to\infty$, leaving only the dependence on $x$ and the
trap size $l$.  We have studied the scaling behaviour of the bipartite
entanglement with increasing the trap size, keeping the other model parameters
fixed at the quantum critical point of the homogeneous system without trap.

As a theoretical laboratory, we have considered the quantum XY chain in an
external space-dependent transverse field, acting as a trap for the spinless
fermions of its quadratic Hamiltonian representation.  This model presents a
quantum critical transition belonging to the 2D Ising universality class, thus
described by a CFT with central charge $c=1/2$.  In the presence of the trap
and for large trap sizes, the quantum critical behaviour can be described in
the framework of the TSS theory~\cite{CV-10}, where the quantum critical
behaviour is cast in the form a TSS, cf.\ (\ref{freee}), with a nontrivial
trap exponent $\theta=p/(p+1)$, which determines how the length scale of the
critical modes at the critical point diverges with increasing trap size, i.e.,
$\xi\sim l^{\theta}$.  Exploiting the quadratic spinless fermion
representation of the XY chain, we have computed the bipartite von Neumann and
R\'enyi entanglement entropies up to $l\approx 10^4$.  Our results show that
at the quantum critical point they behave as
\begin{equation}
\eqalign{{\rm lim}_{L\to\infty} S_\alpha(L/2-x;L) =
C_\alpha \left[ \ln \xi_1 + \ln R_\alpha +
e_\alpha + f_\alpha(X) + \Or(\xi_1^{-1/\alpha})\right],
\\
\xi_1 = a_1 \gamma^{\theta/p} l^{\theta} \left[ 1 + \Or(l^{-\theta})\right],
\quad X\equiv x/\xi_1,
\quad
\xi_\alpha/\xi_1 = R_\alpha + O(\xi_1^{-1/\alpha}), 
} \label{salphasca}
\end{equation}
where $C_\alpha=c(1+\alpha^{-1})/12$, $c=1/2$ is the central charge,
and $e_\alpha$ is the same constant entering the entanglement
entropies of homogeneous systems with open boundary conditions, cf.\
(\ref{egamma}). The entanglement length scales $\xi_\alpha$ are defined by
imposing $S_\alpha(L/2;L)\equiv C_\alpha \left( \ln \xi_\alpha + e_\alpha\right)$; 
the length scales $\xi_\alpha$, and in particular 
$\xi_1$ derived from the von Neumann entropy,
behave consistently with the predictions of the TSS theory, with
$\theta=p/(p+1)$ and $a_1$ independent of $\gamma$. The asymptotic
ratios $R_\alpha$ and the functions $f_{\alpha}(X)$ depend on the
power of the potential, show a small dependence on $\alpha$, but are
universal with respect to the parameter $\gamma$, see figures
\ref{s1xis}, \ref{s2p24}, and \ref{s3p24}.  Moreover, in the
$p\to\infty$ limit they tend to $f_\alpha(X)=\ln\cos \pi X$ with
$X=x/L$, which is the behaviour in the homogeneous system with open
boundary conditions.

We have studied the scaling of the entanglement of confined particle systems
described by the 1D BH model (\ref{bhm}) in the presence of a power-law
confining potential, with the center of the trap in the superfluid phase.
This model is of experimental relevance because it describes cold atoms in
quasi-one-dimensional optical lattices. We have computed bipartite
entanglement in the hard-core limit of the model, which allows very accurate
numerical calculations by exploiting a map into a lattice model of free
spinless fermions.  In the absence of the trap, the continuum theory
corresponding to the gapless superfluid phase is a 2D massless boson field
theory, thus a CFT with $c=1$.  In this region the trap-size dependence
presents new features with respect to the XY chain: a multiscaling behaviour
with the existence of different length scales which diverge with distinct
power laws in the TSS; the presence of level crossings of the lowest states at
finite trap size, which gives rise to periodic modulations of the amplitudes
of the asymptotic power-law behaviours, with a period given by the
asymptotically constant interval between the level crossings.  We have shown
that the bipartite entanglement entropy presents the same features as well.
We have computed the von Neumann and R\'enyi entropies at $\mu=0$
corresponding to half filling $f=1/2$ in the absence of the trap, up to trap
sizes $l\approx 10^3$, corresponding to total particle numbers $N=\Or(10^3)$.
Our results provide a robust numerical evidence of the following scaling
behaviour
\begin{equation}
\eqalign{\fl {\rm lim}_{L\to\infty} S_\alpha(L/2-x;L) = C_\alpha
\Bigl\{\ln\xi_e + e_\alpha + f(X) \\
+\,\xi_e^{-1/\alpha} 
\left[(-1)^x g_{\alpha,o}(Y,\varphi) +  g_{\alpha,s}(Y,\varphi)\right]
+ \Or(\xi^{-2/\alpha})\Bigr\}, \\
\xi_e = a_e l, \qquad X\equiv x/\xi_e, \qquad Y\equiv x/l^{p/(p+1)},
}\label{salphascaxx}
\end{equation}
where $e_\alpha$ and $a_e$ are constant given in equations (\ref{emufunc}) and
(\ref{xiemu0}) respectively.  The leading spatial dependence $f(X)$ turns out
to be independent of $\alpha$; it depends on $p$, approaching the CFT
dependence $f(X)=\ln\cos(\pi X)$ with $X=x/L$ when increasing $p$, which is
the result holding in a homogeneous chain of size $L$.  The
$\Or(\xi_e^{-1/\alpha})$ term presents an even-odd spatial effect proportional
to $(-1)^x=\rme^{2\rmi k_Fx}$, scaling as a function of $Y=x/l^\zeta$ with
$\zeta=p/(p+1)$, modulated by the variable $\varphi$ defined in equation
\eref{barphildef}, see figures \ref{s2lms1}, \ref{s2lms1p4}, \ref{s3lms1}.
For the harmonic potential, an accurate description of
$\Or(\xi_e^{-1/\alpha})$ term is given by $(-1)^N b_\alpha \left[(-1)^x \cos(c
  Y^3) -1\right]$, see figure \ref{s23ms1y3}.

The above scaling behaviours have been mostly obtained by analyzing the scaling
of numerical (practically exact) results at finite trap size.  Of course, it
would be worthwhile to derive analytically the asymptotic trap-size dependence
of the entanglement entropy, which may lead to further physical insights.

\ack
Helpful discussions with Pasquale Calabrese are gratefully acknowledged.

\end{document}